\documentclass{pasa}%

\usepackage{graphicx}

\title[CURVEPOPS I: interacting binaries and type II SNe]{Supernova lightCURVE POPulation Synthesis I: including interacting binaries is key to understanding the diversity of type II supernova lightcurves. }

\author[Eldridge et al.]{J.J. Eldridge$^1$, L. Xiao$^{1,2}$, E.R. Stanway$^{3}$,  N. Rodrigues$^{1}$, N.-Y. Guo$^1$
\affil{$^1$Department of Physics, Private Bag 92019, University of Auckland, Auckland 1010, New Zealand }%
\affil{$^2$CAS Key Laboratory for Research in Galaxies and Cosmology, Department of Astronomy, University of Science and Technology of China, Hefei, 230026, China}
\affil{$^3$Department of Physics, University of Warwick, Gibbet Hill Road, Coventry CV4 7AL, UK}
}%

\jid{PASA}
\doi{10.1017/pas.\the\year.xxx}
\jyear{\the\year}

\usepackage{aas_macros}
\usepackage{hyperref} 
\hypersetup{colorlinks,citecolor=blue,linkcolor=blue,urlcolor=blue}

\begin{document}

\begin{frontmatter}
\maketitle

\begin{abstract}
We present results of a supernova light-curve population synthesis,
predicting the range of possible supernova lightcurves arising from a
population of progenitor stars that include interacting binary
systems. We show that the known diversity of supernova lightcurves can
be interpreted as arising from binary interactions. Given detailed
models of the progenitor stars, we are able to the determine what
parameters within these stars determine the shape of their supernova
lightcurve. The primary factors are the mass of supernova ejecta and
the mass of hydrogen in the final progenitor. We find that there is a
continuum of light-curve behaviour from type IIP, IIL to IIb
supernovae related to the range of hydrogen and ejecta masses. Most
type IIb supernovae arise from a relatively narrow range of initial
masses from 10 to 15\,M$_{\odot}$. We also find a few distinct
lightcurves that are the result of stellar mergers.
\end{abstract}

\end{frontmatter}

\section{INTRODUCTION }
\label{sec:intro}

Core-collapse supernovae (SNe) are the explosive events that mark the
death of a massive star. Since their origin was identified by
\citet{1934PNAS...20..254B}, many thousands of supernovae have been identified
and observed. In recent times,  large programs have begun to systematically search for
them in nearby and distant galaxies
\citep[e.g.][]{2009PASP..121.1395L,2011MNRAS.412.1441L}. With the resultant large samples it becomes possible to start to look at the trends and diversity expected from these events.

Since soon after their discovery, supernovae have been split into categories determined
by their observational properties. The first split is made by studying
their spectra to determine their composition
\citep[see][]{1997ARA&A..35..309F}. If no hydrogen is observed an event
is classified as type I, while those with hydrogen are type II.  Type I events include the
thermonuclear detonations of white dwarfs (type Ia) and the
core-collapse of massive stars that have lost their hydrogen
envelopes (types Ib and Ic). We will not consider type I supernovae here.

In this article we concentrate on the more common, hydrogen-rich type II
core-collapse supernovae. These are divided into sub-classes \citep{1997ARA&A..35..309F} first by their photometric behaviour; i.e. whether there is a constant luminosity plateau (type IIP) or the lightcurve declines linearly (type IIL). Analysis of spectroscopic behavior yields a further two subclasses. Type IIb supernovae appear as type II events (with hydrogen)
in the first few weeks of observations, but then the hydrogen lines disappear from the
spectra and the spectrum resembles instead a type Ib SN. Type IIn
supernovae instead have narrow lines of hydrogen indicating low expansion
velocities inconsistent with the broad lines seen in other type II events, although they normally also show some features of other supernova classes. The canonical explanation for this last subclass is that the narrow lines arise through interactions of their expanding shocks with a circumstellar medium (perhaps populated by ejecta from earlier phases of the progenitor star). We do
not consider them here as their characteristics are related only indirectly to the star that gives rise to them. 

A final supernova subclass that must be considered before moving on are the type II-peculiar or
SN\,1987A-like SNe. SN\,1987A occurred in the Large Magellanic Cloud and
was the first supernova to have an identified progenitor
 in pre-explosion imaging \citep{1987ApJ...321L..41W}. However it had a dim and unusual
lightcurve. This was due to the small size of the progenitor at
explosion, likely because it was the result of a stellar merger
\citep{1992PASP..104..717P}. Similar events are rare but several other examples
have been found since the SN\,1987A archetype was identified \citep[e.g.][]{2012A&A...537A.141P}.

Over the past few years several groups have collated large samples of type II lightcurves, attempting to determine the true population parameters for this class of explosion \citep[e.g.][]{2012ApJ...756L..30A,2014ApJ...786...67A,2014MNRAS.442..844F,2014MNRAS.445..554F,2015ApJ...799..208S,2016MNRAS.459.3939V,2017ApJ...841..127M,2017ApJS..233....6H}.  Type IIP supernovae are the most common but there is some debate as to whether there is a continuum of behaviour between the IIP, IIL and IIb sub-classes or whether these are entirely distinct. One problem is that the completeness of individual observed lightcurves has to be considered. Poor sampling of a lightcurve may lead to poorly constrained plateau phases or failure to identify spectral time evolution (as in the case, for instance, of the type IIb to Ib transition). In the past many supernovae were not identified until relatively late (i.e. a few days to weeks past peak luminosity), and information was missing on their early time behaviour. However,  supernovae are more likely to be found early with the current generation of automated surveys \citep[e.g.][]{2009PASP..121.1395L,2012AAS...22043203S} which also observe sources over an extended period with a well-defined cadence.

One method to attempt to understand supernovae is to model the explosions and attempt to link the observed supernovae to stellar progenitors. This an established field with many groups and codes attempting to gain insight into supernovae with this method \citep[e.g.][]{1994A&A...281L..89U,2005AstL...31..806U,2007A&A...461..233U,2008A&A...491..507U,2009A&A...506..829U,2010MNRAS.408..827D,2011MNRAS.414.2985D,2011ApJ...729...61B,2012ApJ...757...31B,2014MNRAS.440.1856D,2014AJ....148...68B,2015ApJ...814...63M,2016MNRAS.458.1618D,2016ApJ...829..109M,2017ApJ...838...28M}. 

Insights can also be achieved by investigating the relative rates of different supernova types, especially the rate of type Ib/c to type II, it has become clear that interacting binary stars are an important factor in creating this difference \citep{1992ApJ...391..246P,1998A&A...333..557D,2008MNRAS.384.1109E,2011MNRAS.412.1441L,2011MNRAS.412.1522S,2013MNRAS.436..774E,2015MNRAS.452.2597X,2017ApJ...837..120G}. The abscence or prescence of hydrogen in stellar progenitor models can be easily determined. However determining which subtype of hydrogen-rich supernovae a model may produce is more difficult. Typically a mass of hydrogen in the envelope is assumed but these masses are typically an informed guess or an estimate calibrated from supernova models \citep{2010MNRAS.408..827D,2011MNRAS.414.2985D}.

In this paper we present the first step in combining these methods in the supernova lightCURVE POPulation Synthesis project, CURVEPOPS. Our novel approach is to create a synthetic population of supernova lightcurves from an already well-established and thoroughly tested population of synthetic stellar models. We use the BPASS version 2 stellar models. BPASS, the Binary Population and Spectral Synthesis project \citep{2017PASA...34...58E} uses a custom set of detailed stellar evolution models to make predictions about a large range of parameters that can be observed for stellar systems. It has been validated and compared to a wide range of observed supernovae \citep[e.g.][]{2015ApJ...814...63M,2016ApJ...829..109M,2017ApJ...838...28M,2016ApJ...826...96P,2017ApJ...851..138D,2018ApJ...858...15M} to confirm that the stellar models reflect the evolution of real stars. Crucially the model grid includes the effect of binary interaction and mass transfer on the structure and properties of each star.
We select a small but representative number of our stellar models that are expected to undergo a core-collapse supernova and simulate this death throw for each star using the publicly-available Supernova Explosion Code \citep[SNEC, ][]{2017ApJ...838...28M}. 
We then compare the range of resulting lightcurves if only single star progenitors are used, and then if the full range of available binary stellar progenitor models are used. We are thus able to determine whether, as suggested by  \citet{1995NYASA.759..360N,1996IAUS..165..119N},  binary stars are responsible for the diversity of supernova types. The advantage of this method is that we will not only be able to classify supernova types similar to those observed in the Universe, but we may also identify samples of supernovae that may be overlooked by observational campaigns that are not optimised to detect these faint or fast (or both) events.

The outline of this paper is as follows: we first briefly summarize the BPASS stellar models and how we selected those we use here. We describe how these are input into SNEC and outline our explosion parameters. We then discuss the resulting population of supernova lightcurves expected from single star and binary populations, before listing caveats to our study and discussing our conclusions.

\section{Method}


\subsection{Stellar Models}

We use the v2 stellar models used in the Binary Population and Spectral
Synthesis code {BPASS} \citep{2017PASA...34...58E}. These are the base stellar evolution models used by all the version 2 releases of the BPASS population synthesis (to date v2.0, v2.1 and v2.2). The models
are calculated using a modified version of the Cambridge STARS code that has been adapted to follow binary evolution \citep[see][for full details]{2017PASA...34...58E}. The individual stellar models are available from the website \texttt{bpass.auckland.ac.nz} at moderate time resolution, but these are derived from much larger evolution code outputs with finer time steps, sufficient to trace the effects of binary mass transfer and mergers on stellar evolution. 

The models are fully described elsewhere but we summarise the most important details
here. The models are calculated from the zero-age main-sequence up to
the end of core carbon burning. This is close enough to the time of
core-collapse that the parameters of the star at explosion (a necessary input for our lightcurve modelling)
will not vary significantly. For this study we use models with a metallicity mass fraction of
$Z=0.014$ which is close to the estimated metallicity of massive stars in the Solar neighborhood \citep{2012A&A...539A.143N}. The full range of stellar
initial masses spans from $0.1M_{\odot}$ to $300M_{\odot}$, with a range of initial binary component mass ratios from $q=0.1$ to
$0.9$ in 0.1 steps and a range of initial periods from
$\log (a/R_{\odot})=1$ to 4 in 0.2 dex steps. 

We stress that this group of binary models are unique because they are
calculated in a detailed stellar evolution code rather than by rapid
population synthesis \citep[e.g.][]{2004MNRAS.348.1215I}. As discussed
in \citet{2008MNRAS.384.1109E} and \citet{2017PASA...34...58E}, the results of mass loss and
binary interactions are followed in greater detail. This means that  rather than rather than stripping hydrogen envelopes entirely as a result of binary interactions, we are able to calculate the atmosphere parameters based on the stellar structure and follow the stripping progress. This is a more physically motivated treatment; we find that binary
interactions remove material until the envelope contracts within the
star's Roche Lobe. After this point stellar winds are required to
remove the remaining hydrogen - a process which may or may not occur. When
comparing the results of rapid binary evolution and detailed binary
models, \citet{2008MNRAS.384.1109E} found this to be the primary
difference between the two methods. For determining the supernova type it is
vital to correctly predict the amount of hydrogen left on the progenitor star's surface and
so detailed models must be used.

From this large model set we identify those models that end their evolution having completed core carbon burning and have a carbon-oxygen core mass greater than 1.38$M_{\odot}$. Out of our approximately 250,000 stellar models over all metallicites we find that 100,000 met these criteria for a core-collapse supernovae. To make the supernova synthesis a more tractable problem, we only consider the type II, hydrogen-rich supernovae here, and so we require that the surface hydrogen mass fraction is $X>10^{-3}$ and that the final hydrogen mass in the star is $>10^{-3}$M$_{\odot}$. In addition we only consider the explosion from the primary stars and restrict our analysis to a limited number of initial masses drawn from our full grid. These are 5, 6, 7, 8, 9, 10, 12, 15, 20, 25, 30 and 40$M_{\odot}$.  These cuts yield 637 individual supernova progenitor stellar structures from our binary models. We also consider the same 12 initial masses from our single star models. For each of these models we extract the relevant interior structure and composition details and prepare these as input to SNEC.

\subsection{Exploding BPASS stellar models in SNEC}

\begin{figure*}
\begin{center}
\includegraphics[width=168mm]{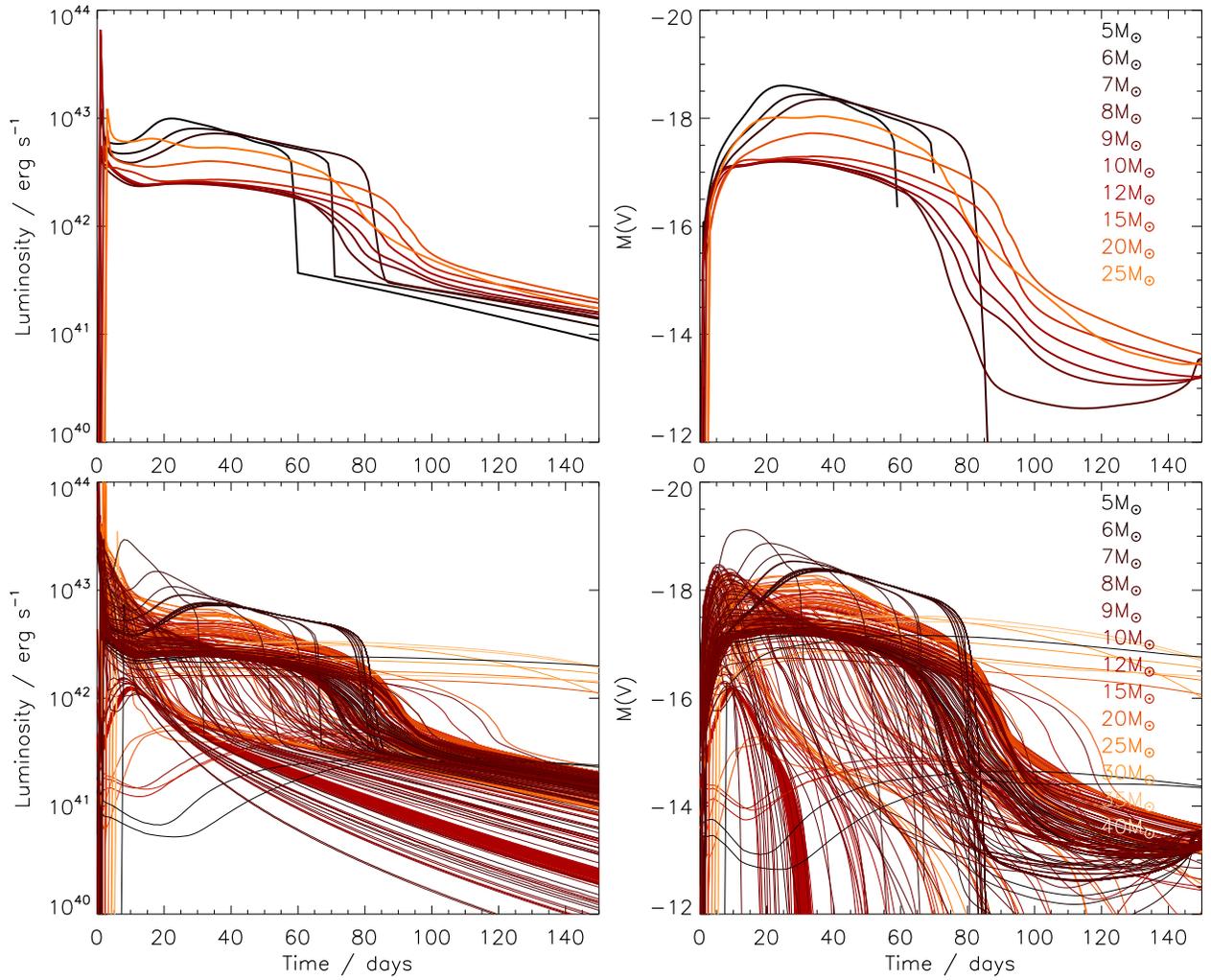}
\caption{The bolometric (left panels) and visual magnitude (right panels) lightcurves of all our type II supernova models. The upper panels are for single stars and the lower panels are for binary models with the same initial primary masses. Increased diversity in lightcurve behaviour is evident due to binary interactions.}\label{Fig1}
\end{center}
\end{figure*}

The SuperNova Explosion Code, SNEC, is an open source modelling code that is available online from \texttt{https://stellarcollapse.org/SNEC} and which has been appied to modelling a number of aspects of supernovae \citep[e.g.][]{2015ApJ...814...63M,2016ApJ...829..109M,2017ApJ...838...28M}. We have reformatted our stellar models so that they can be input into SNEC and exploded. We only include the composition variables that we have within the BPASS stellar evolution code. These are hydrogen, helium, carbon, nitrogen, oxygen, neon, magnesium, silicon and iron abundances. We specify the nickel-56 variable within the explosion parameters of the SNEC model, rather than the input stellar structure. We make all these input files available on the PASA datastore as well as on the BPASS website (\texttt{http://bpass.auckland.ac.nz}).

Within SNEC, certain other parameters must be specified in addition to the progenitor structure. These are the explosion energy, nickel mass, amount of nickel mixing and excised mass. Each of these are somewhat dependent on the stellar progenitor. For simplicity in this initial analysis, we have kept a consistent prescription for all these explosion parameters over our grid, independent of input stellar structure. The values we select are as follows,
\begin{enumerate}
\item The explosion energy is assumed to be $10^{51}\ {\rm erg \, s^{-1}}$ and is input as a thermal bomb.
\item The nickel mass is taken to be 0.05$\,$M$_{\odot}$.
\item The excised remnant mass, $M_{\rm excised}$, is taken to be the estimated remnant mass in the BPASS models as described in \citet{2004MNRAS.353...87E}. This is determined by first calculating the mass that can be ejected during the explosion as the mass fraction of the progenitor with a gravitational binding energy less than $10^{51} {\rm erg \, s^{-1}}$. Any remaining mass after this is removed is designated as the remnant mass. We note that models with a remnant mass less than 1.38\,M$_{\odot}$ are not considered to explode as supernova due to evolutionary pathways onto the Asymptotic Giant Branch which lead instead to collapse to a white dwarf. 
\item The nickel-56 manufactured by explosive nucleosynthesis during the supernova is generated primarily at the stellar core, but can be rapidly mixed out into the expanding material, which will ultimately be ejected. 
The nickel boundary mass, which is the integrated mass fraction interior to the outermost layer which contains nickel, is calculated as $M_{\rm Ni\,bounary}=M_{\rm total} - 0.1M_{\rm ejecta}$, where $M_{\rm total}$ is the progenitor's total mass and $M_{\rm ejecta}$=$M_{\rm total} - M_{\rm excised}$. This therefore assumes significant mixing of nickel into the envelope. 
\end{enumerate}


We show the results of the full simulation set in Figure \ref{Fig1}. It is immediately apparent that a much greater variety of lightcurves are possible from  binary progenitors than from single stars. The latter supernova types all appear to have long plateaus and thus to be of type IIP. We note that the lower mass single stars (initial masses of 5, 6 and 7\,M$_{\odot}$) in reality are unlikely to explode in a core-collapse supernova but are included for comparison to the binary progenitors with these initial masses that do explode. 

Several of the tracks rapidly decrease in V band magnitude while the bolometric luminosity for the same supernova model does not show this behaviour. This is likely a computational artifact that demonstrates a limitation of the SNEC code.  SNEC  uses a black body approximation to calculate the broad-band magnitudes that would be observed. This approximation breaks down when more than 5\% of the total luminosity arises from nickel-56 decay beyond the photosphere. Thus for many of our lightcurves with small ejecta masses, their simulated broadband magnitudes drop away although the bolometric luminosities continue to hold.

\subsection{Analysing the explosion models}

A brief inspection of Figure \ref{Fig1} demonstates that the known principle types of type II supernovae are represented in this population, namely lightcurves that appear to be like IIP, IIL and IIb events. Buried in this population there may also be other lightcurves that do not fit neatly into these categories. We have therefore taken two approaches to attempt to understand the variety in our supernova models. 

Firstly we compare the lightcurves from progenitor stars with the same initial mass to gain some understanding of how much a role this plays in determining a supernova type.

Second we have visually inspected that 637 supernova lightcurves and identified groups which demonstrate similar evolution in luminosity. While an automatic fitting alogorithm could be used, it is instructive here to approach the problem in the same way observers first used to type different supernovae. We expect to find IIP, IIL and IIb lightcurves but also consider very different lightcurves that do not fit into any of these groups. From this typing we are then able to investigate the parameters such as hydrogen mass and radius of the progenitor models that produced each lightcurve. While for observed supernovae this is somewhat model-dependent guess work, here we know the relevant parameters of each of our progenitor models and are able to thus test how the lightcurve shape depends on the nature of the star that exploded.

\section{Results}

\subsection{Comparing stars of similar initial mass}

\begin{figure*}
\begin{center}
\includegraphics[width=\columnwidth]{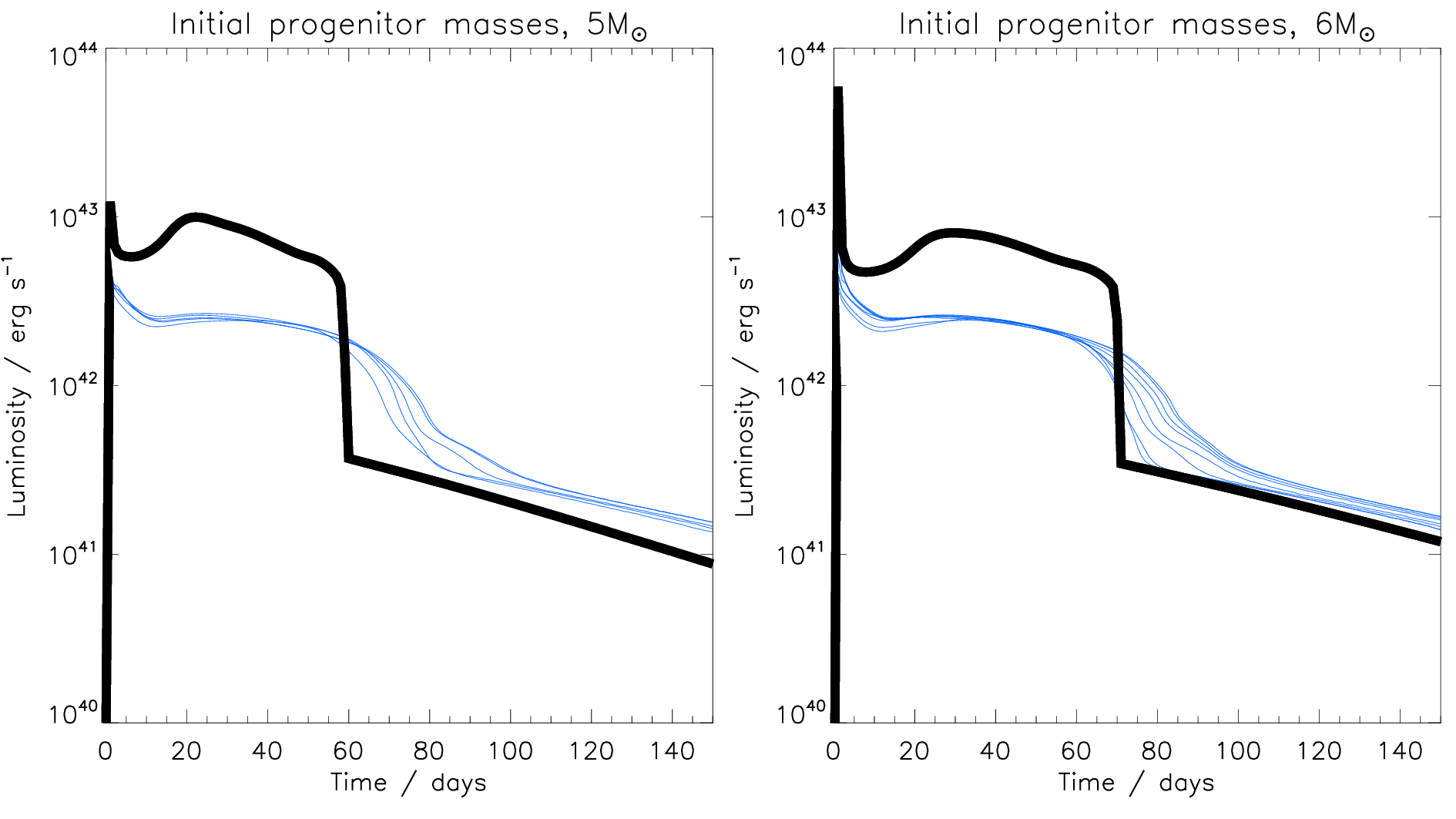}
\includegraphics[width=\columnwidth]{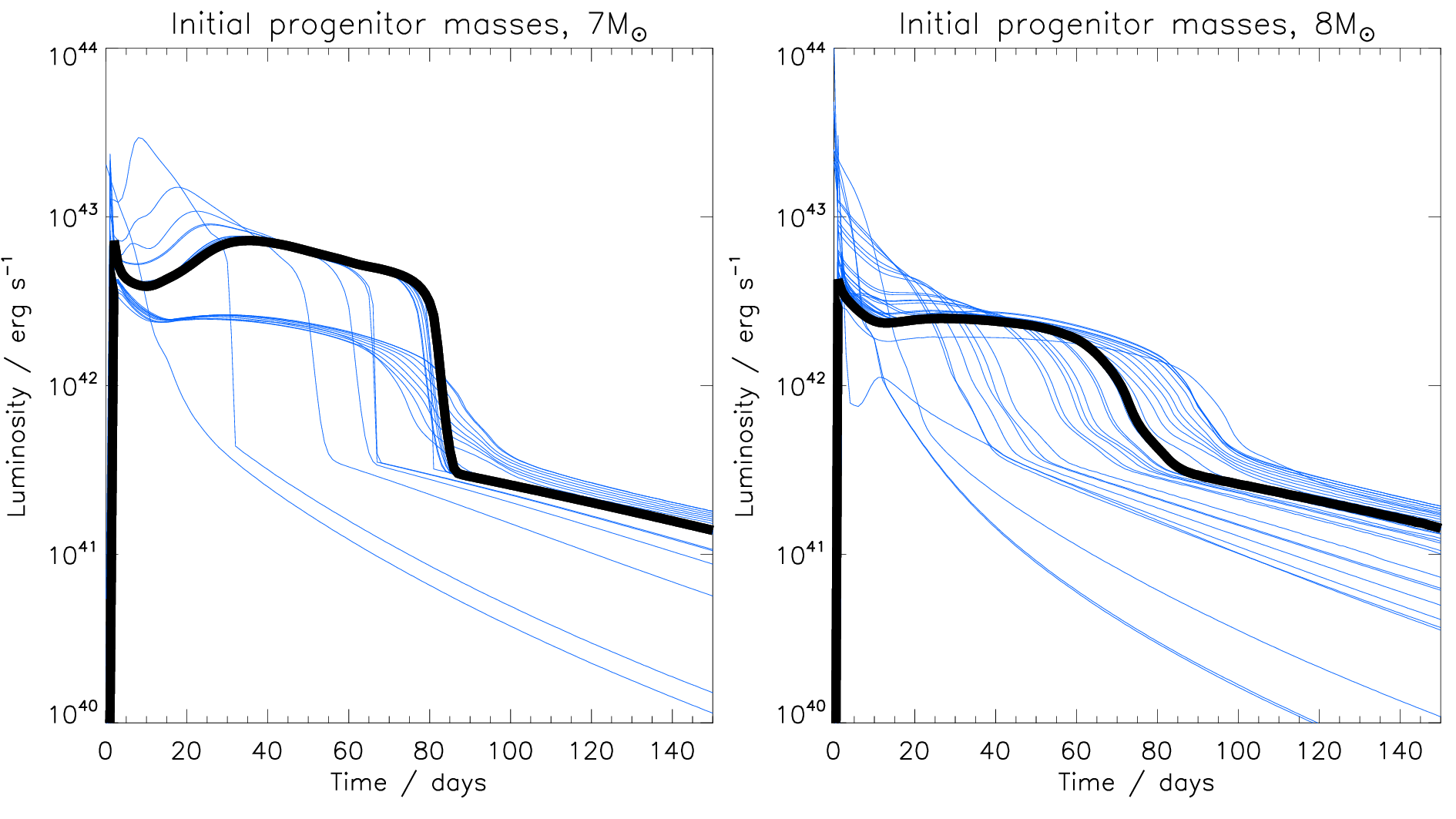}
\includegraphics[width=\columnwidth]{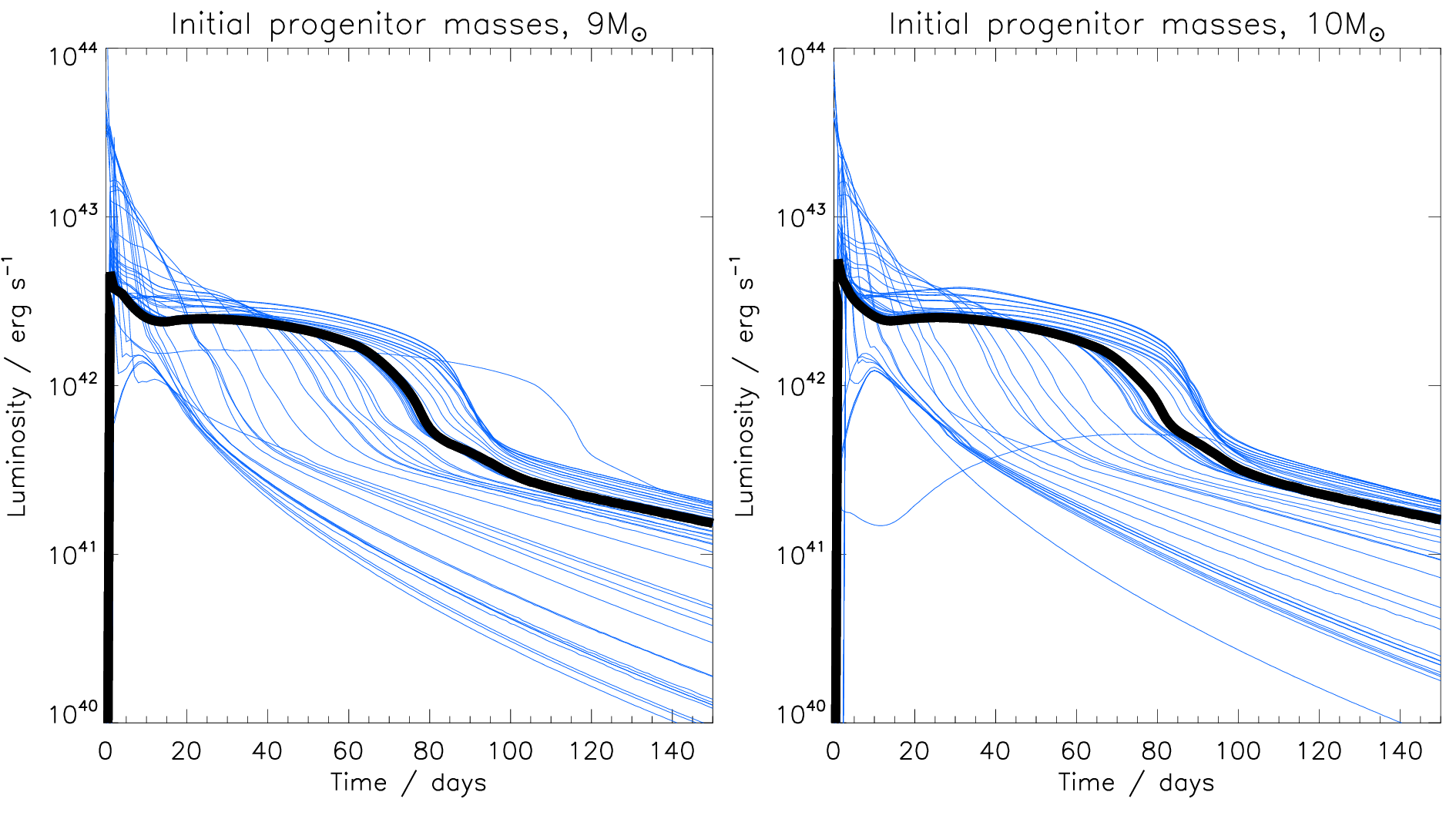}
\includegraphics[width=\columnwidth]{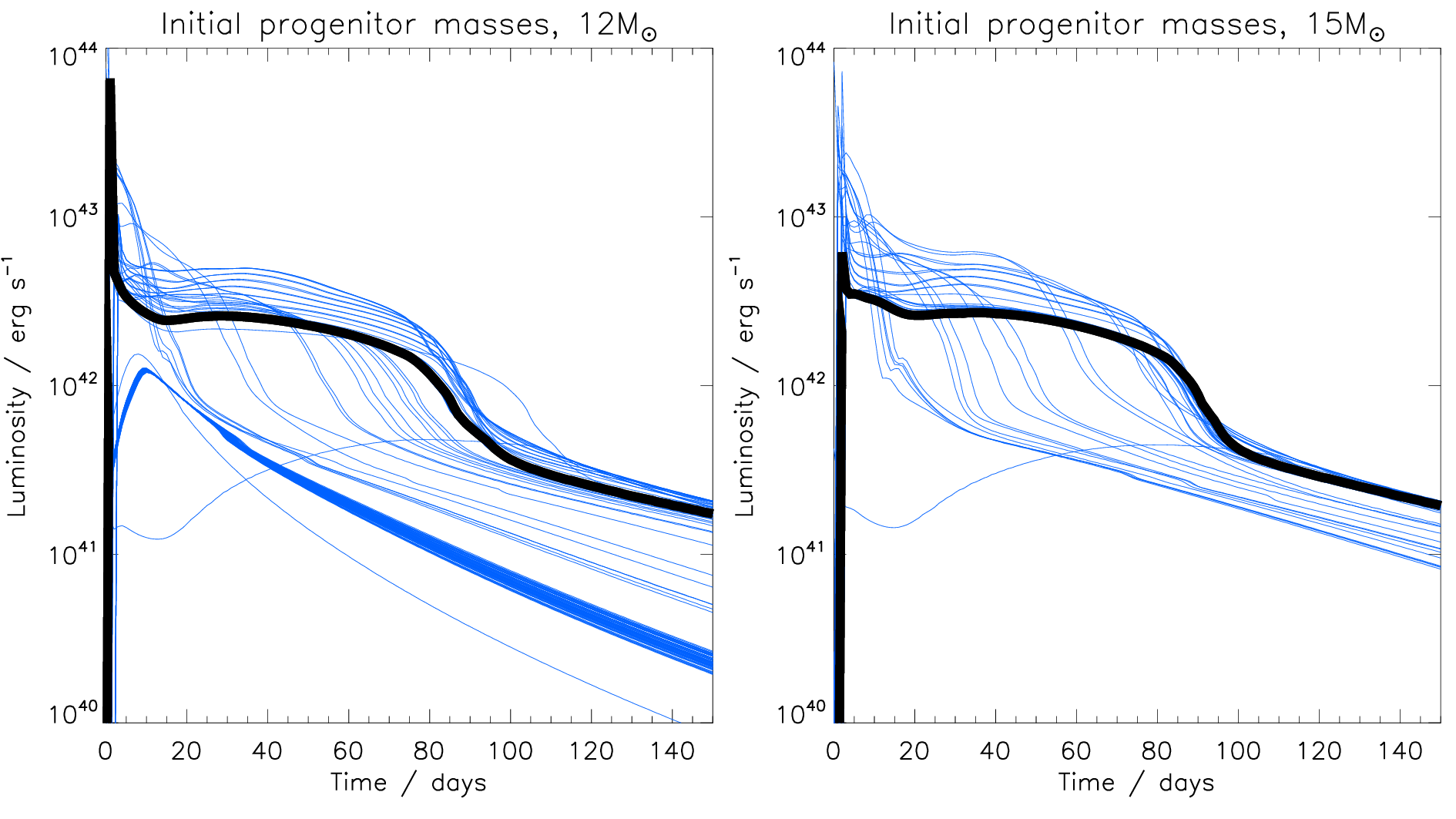}
\includegraphics[width=\columnwidth]{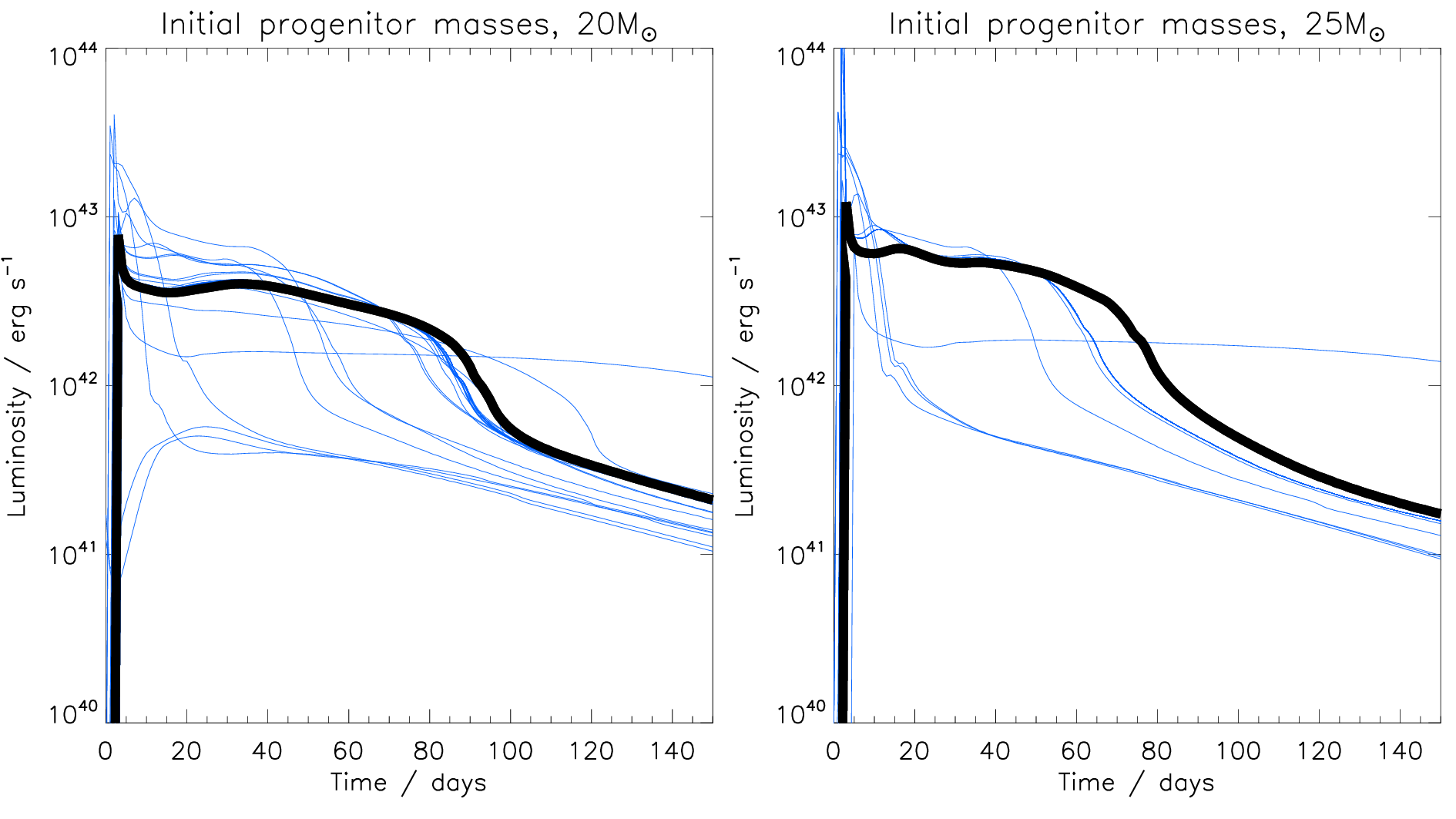}
\includegraphics[width=\columnwidth]{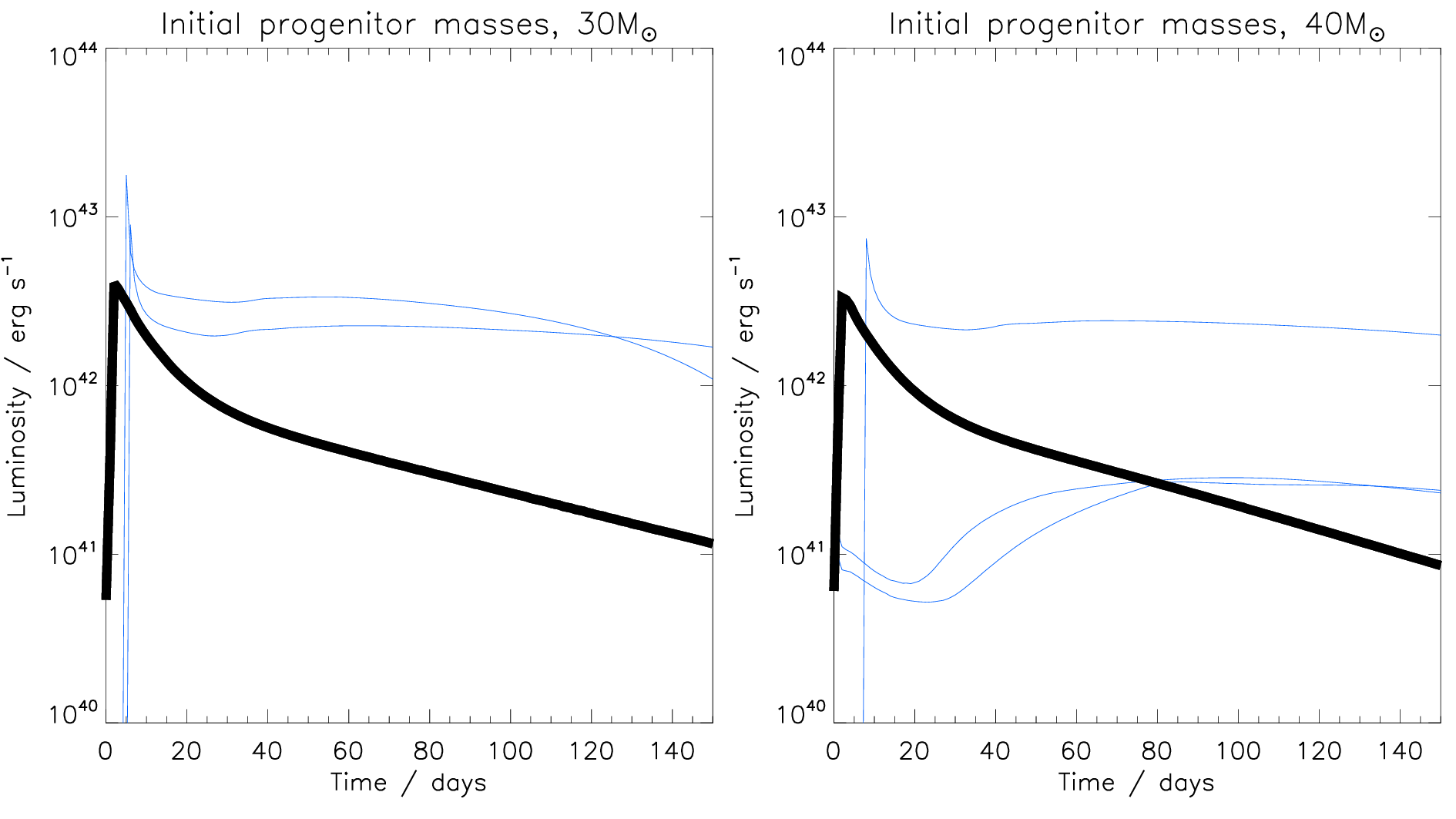}
\caption{Similar to the bolometric lightcurves shown in Figure 1 but here with the models separated by initial mass. The thick black line represents the single star model with that initial mass while the purple thin line represents the lightcurves from interacting binary stars with the same initial mass. }\label{Fig2}
\end{center}
\end{figure*}

In Figure \ref{Fig2} we collect our model lightcurves together by the progenitor initial mass. A number of important features are apparent. For the 5, 6 and 7\,M$_{\odot}$ mass single stars the lightcurves are quite distinct to all the other lightcurves. This is due to their unique structure being AGB stars at the point where the model ends. For these progenitors the core of the star is effectively either a carbon-oxygen or oxygen-neon white dwarf surrounded by a hydrogen envelope. 
These compact cores and low mass hydrogen envelopes give rise to the short and bright plateaus. The more typical plateau-like lightcurves from stars of the same mass are thus the result of mergers. These look like the single-star lightcurves of the 8\,M$_{\odot}$ progenitor. 

In more massive progenitors the inner cores are covered by thicker mantles of helium and thus the cores are less compact.
For the initially 7\,M$_{\odot}$ progenitors we begin to see some lightcurve diversity. This is first apparent in the range of plateau lengths, with one lightcurve not having a plateau at all and being more of a linear shape. This is due to varying amounts of enhanced mass loss due to binary interactions, which in turn causes the ejecta mass to vary. This behaviour continues for more massive stars up to about 20\,M$_{\odot}$. 
At 20\,M$_{\odot}$ there are a range of plateau lengths. The longer plateaus indicate a larger ejecta mass and so indicate that some progenitors gain mass in a merger. The plateaus significantly shorter than a single star plateau at the same initial mass must have lost mass through Roche lobe overflow or stellar winds. The lowest luminosity lightcurve at this mass clearly resembles that of a type IIb event.

The 10, 12 and 15\,M$_{\odot}$ progenitors also exhibit examples with a much fainter and broader lightcurve shape which we describe in the next section, while the more massive progenitors of 20 and 25\,M$_{\odot}$, show either a plateau with a much smaller variety of lightcurves along with cases with a very long apparent plateau of over 150 days.

From this initial analysis it is apparent that most known type II supernova lightcurves arise from stars with initial masses between 8 to 15\,M$_{\odot}$, while masses above and below this range may possibly contribute rarer odd types of type II SNe.

\subsection{Comparing stars with similar lightcurves}

\begin{figure*}
\begin{center}
\includegraphics[width=2\columnwidth]{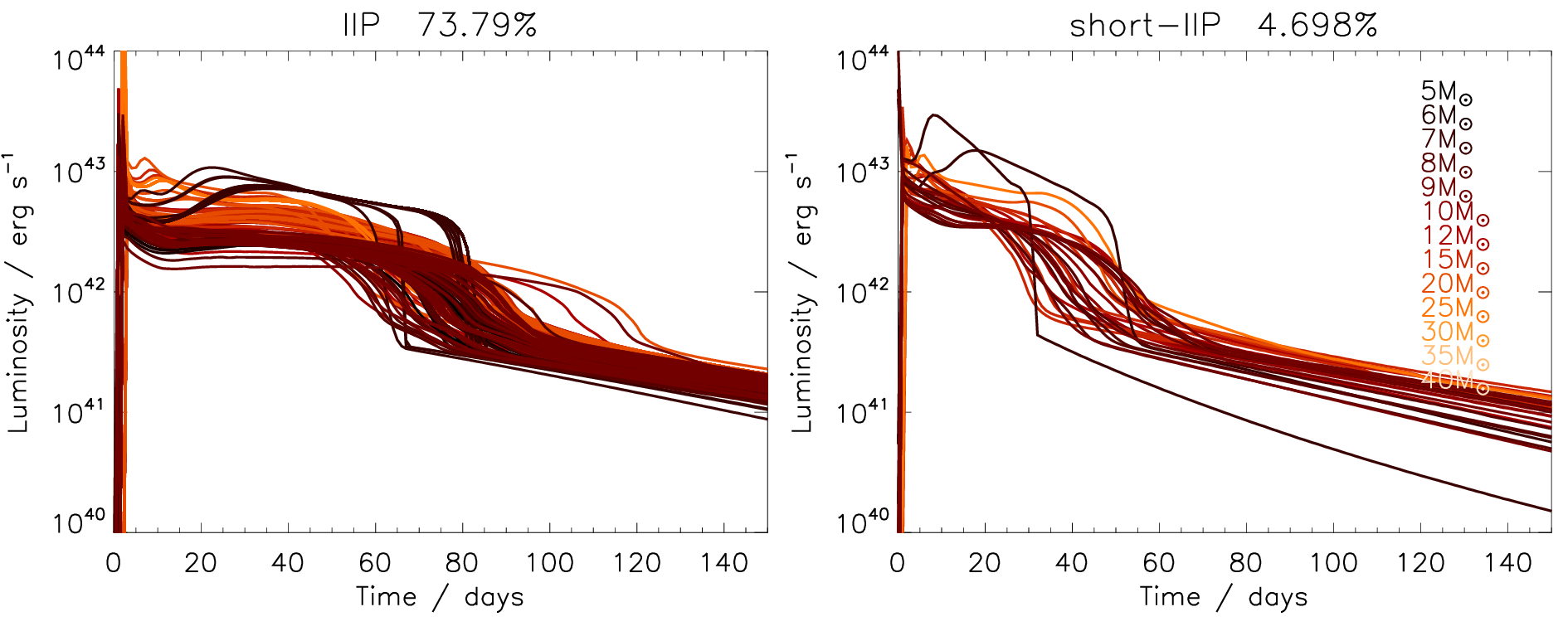}
\includegraphics[width=2\columnwidth]{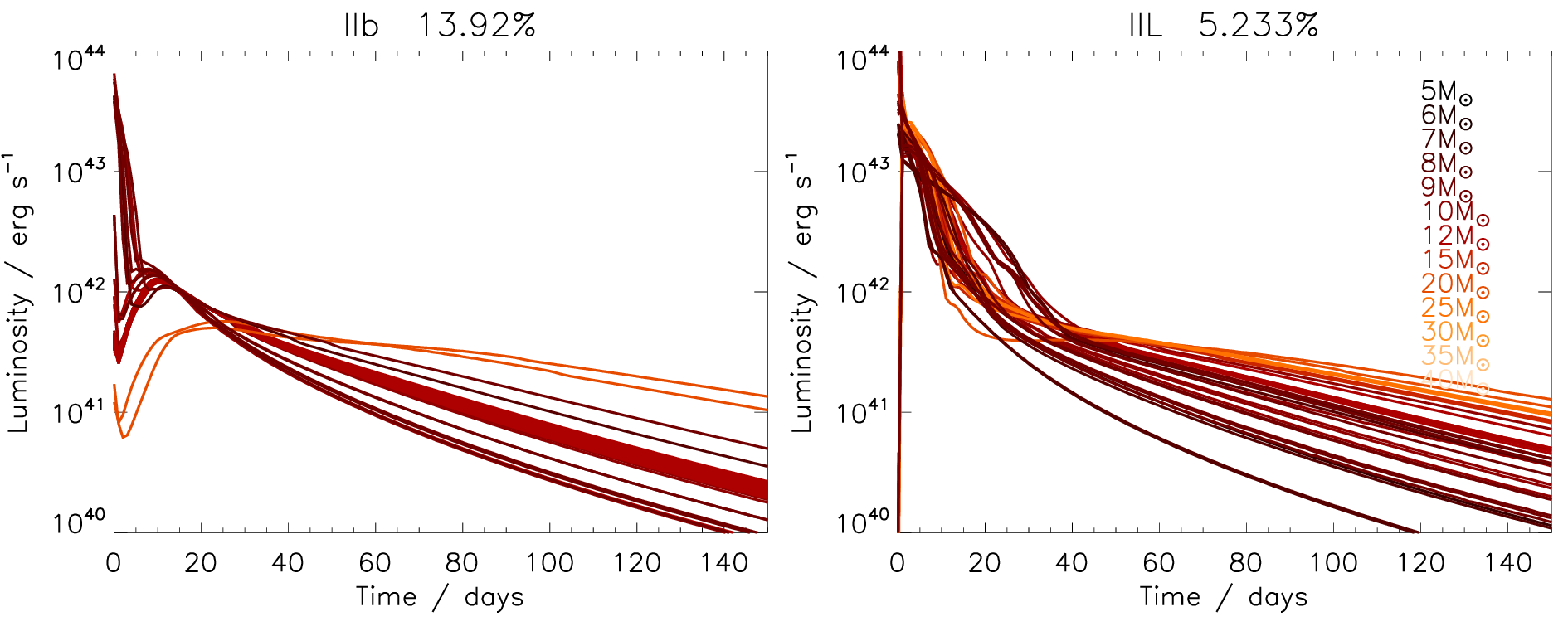}
\includegraphics[width=2\columnwidth]{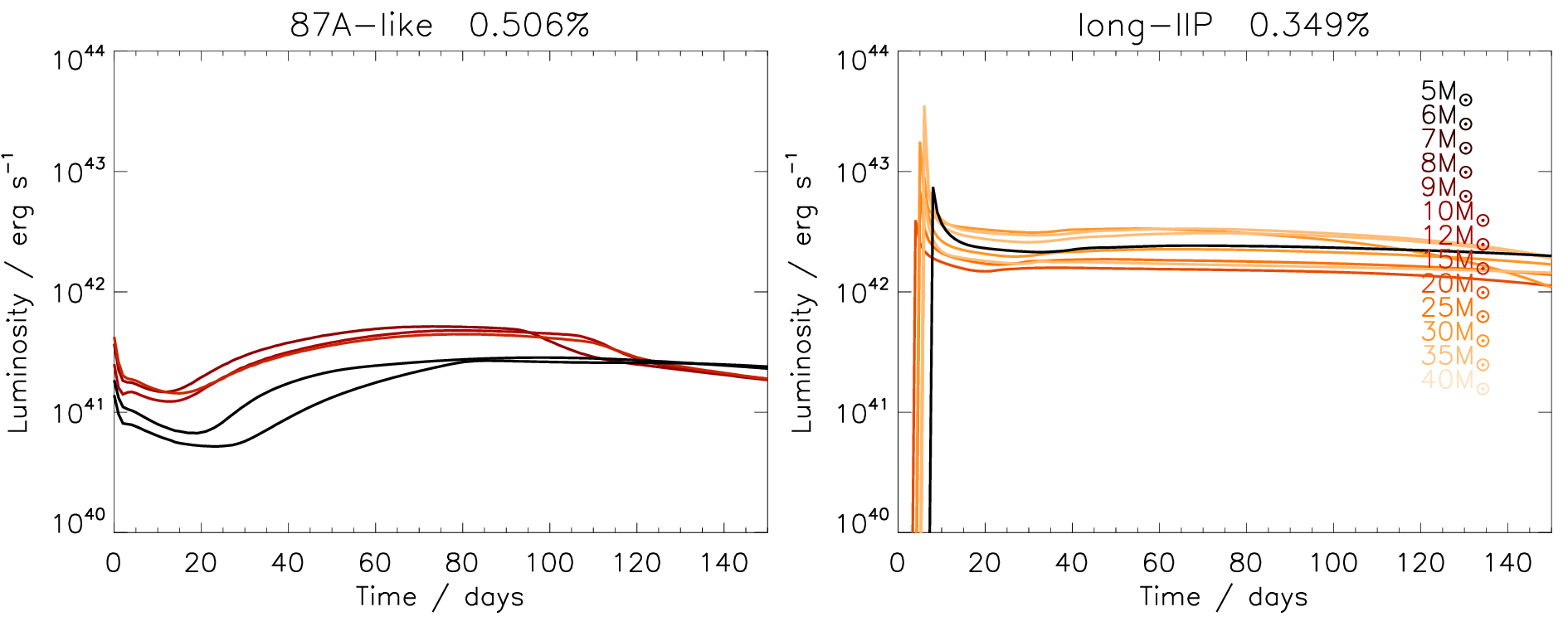}
\caption{The bolometric lightcurves for our synthetic supernova lightcruves separated by eye into various types analogous to observed supernova types. The approximate equivalent supernova type is given in the title of each panel as is the fraction of the models included in that type as a percentage of the total.}\label{Fig3}
\end{center}
\end{figure*}

\begin{figure*}
\begin{center}
\includegraphics[width=2\columnwidth]{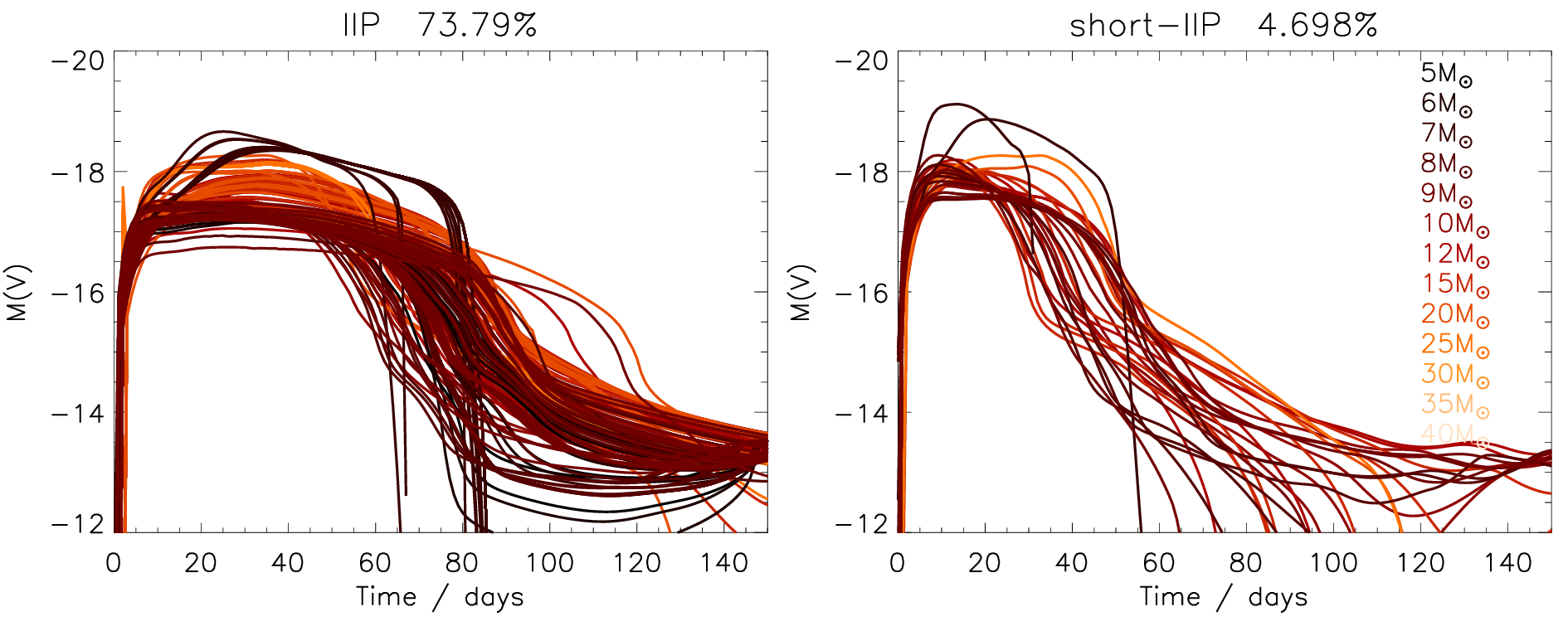}
\includegraphics[width=2\columnwidth]{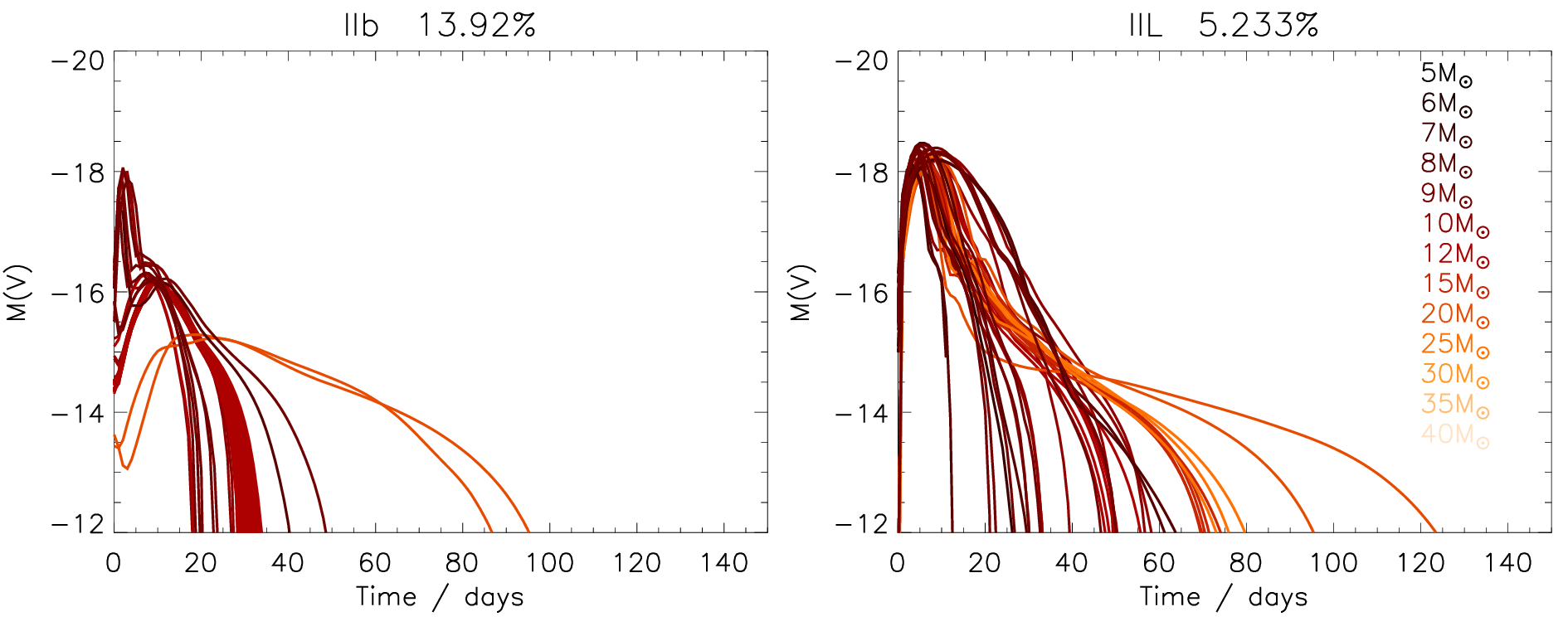}
\includegraphics[width=2\columnwidth]{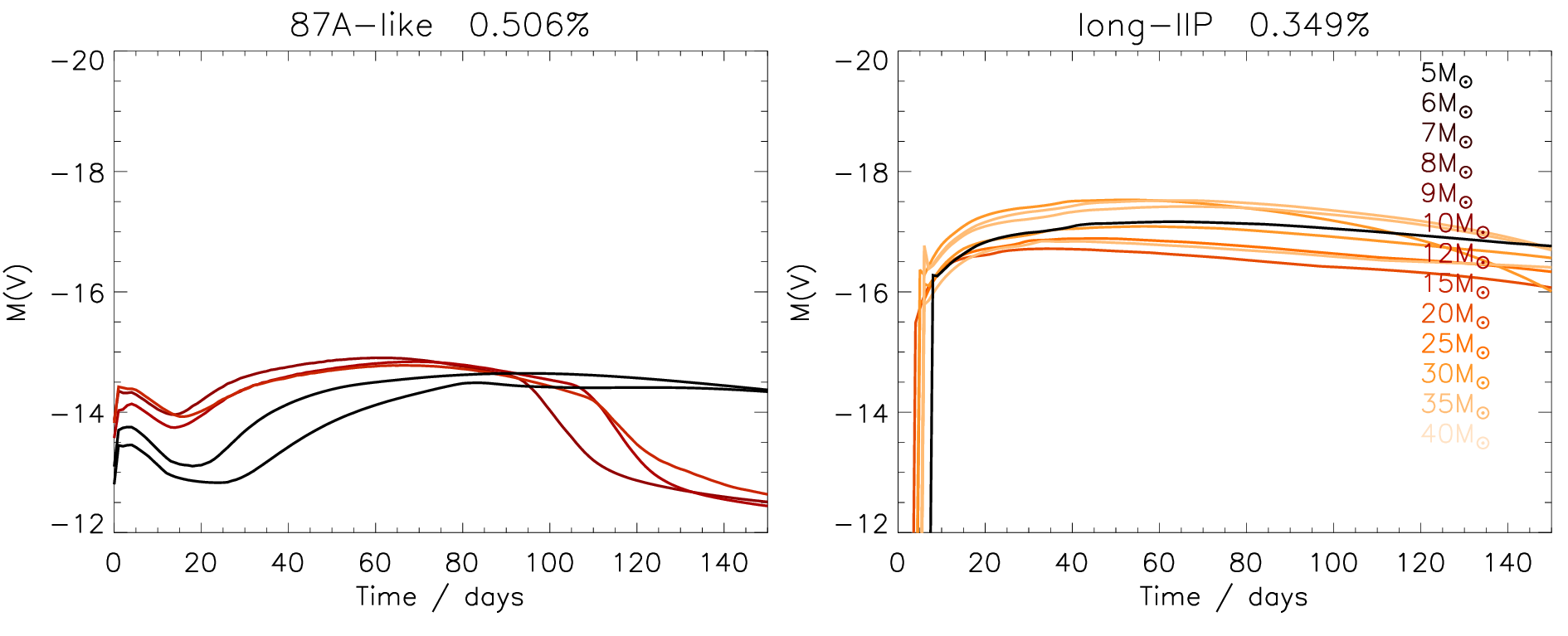}
\caption{The visual magnitude lightcurves for our synthetic supernova lightcruves separated by eye into various types analogous to observed supernova types. The approximate equivalent supernova type is given in the title of the bolometric panel as is the fraction of the models included in that type as a percentage of the total.}\label{Fig3b}
\end{center}
\end{figure*}

An interesting way to analyse our results is to reproduce the empirical classification observers have applied to supernovae over the last century. As discussed in section \ref{sec:intro}, the variety of observed stellar explosions have been classed by the presence or absence of hydrogen (type I/II) and by lightcurve shape (e.g. IIP, IIL, IIb etc). 

Here we undertake an empirical classification of each lightcurve through visual inspection as
type IIP events that have a plateau to their lightcurve, IIL that have a linear decay to their lightcurves and IIb which typically have lightcurves similar to type I supernovae but must have hydrogen in their spectra due to their surface hydrogen abundance at explosion. We also identify 87A-like supernovae. Supernova 1987A exploded in the Large Magallenic Cloud and it the closest and most intensively studied event ever \citep{1987ApJ...321L..41W,2016ARA&A..54...19M}. As the progenitor was a blue supergiant, rather than a red supergiant, the lightcurve was fainter with a unique shape.

When we typed the supernova by eye it was unavoidable to be guided by these types and thus as shown in Figure \ref{Fig3} we have used them as a starting point. However we found the cut off between IIP and IIL was reasonably arbitrary, although the plateau does significantly change shape if the length was less than approximately 40 to 60 days. Similarily the difference between IIL and IIb was difficult to discern. In the end we decided to define the categories type IIP, short-IIP, IIL and IIb, noting again that the classification of borderline lightcurves is ambiguous.

Given the arbitrariness of allocating the supernova types in some cases, we should evaluate how important the exact dividing line is. Population statistics give some insight here. If we use a Salpeter initial mass function and assume that all models with the same initial mass are equally likely (i.e. assuming a flat distribution in $\log$-Period and mass ratio) then we can estimate the relative rate of our determined supernova types. 

We note that these should be interpreted as indicative rather than absolute fractions, given that we are considering the explosion of primary stars alone. We are also considering uniform distributions in period and mass ratio, while recent studies have suggested that massive stars are more likely to be in close, and unequal, binaries than lower mass stars \citep{2017ApJS..230...15M}. Given the narrow mass range considered here, and the other assumptions, we defer a more refined binary parameter weighting to future work.
We have also neglected the circumstellar medium in calculating the lightcurves (thus eliminating likely IIn events) nor varied the metallicity of our progenitors which are likely to have an effect on our predicted numbers. However our initial population synthesis should nonetheless demonstrate the trend and magnitude of subclass fractions in a representative population.

We find that 65.1\% of events are type IIP, short-IIP are 4.1\%, IIL are 4.6\% and IIb are 24.0\% (the small remaining 2.2\% were models that failed to complete). 
From this we  conclude that most of the supernovae are of type IIP and IIb, those of uncertain classification lying between are only a small fraction of the total number of events. Comparison to observed relative rates shows some agreement, \citet{2013MNRAS.436..774E} suggests that IIPs represent 77.5\% of the type II supernova population, IIL are 4.2\% and IIb are 16.9\%. While \citet{2011MNRAS.412.1441L} suggest that IIPs are 70\%, IIL are 10\%, IIb are 12\% and IIn are 9\%. Therefore our simple population synthesis produces a comparable trend in subtypes as both samples.

Finally in Figure \ref{Fig3} there are two types of lightcurves that did not fit into these regular common types. The first were 87A-like lightcurves which represent 0.4\% of the the model population (compared to 1.4\% in observations). These were faint events that took a long time to rise and were similar to the lightcurve of supernova 1987A. The second were long-IIP lightcurves. They had the plateau feature expected for IIPs but with very long plateaus extending beyond 150 days. No such events have been recognised in observational surveys. These were also rare at only 0.3\% of the total supernovae but quite distinct. We suggest that our assumption of constant explosion parameters means is incorrect for these progenitors. They are more likely to have be more or less energetic explosions and thus would appear significantly different to what we predict here.

\subsection{Understanding the progenitor population}

\begin{figure*}
\begin{center}
\includegraphics[width=\columnwidth]{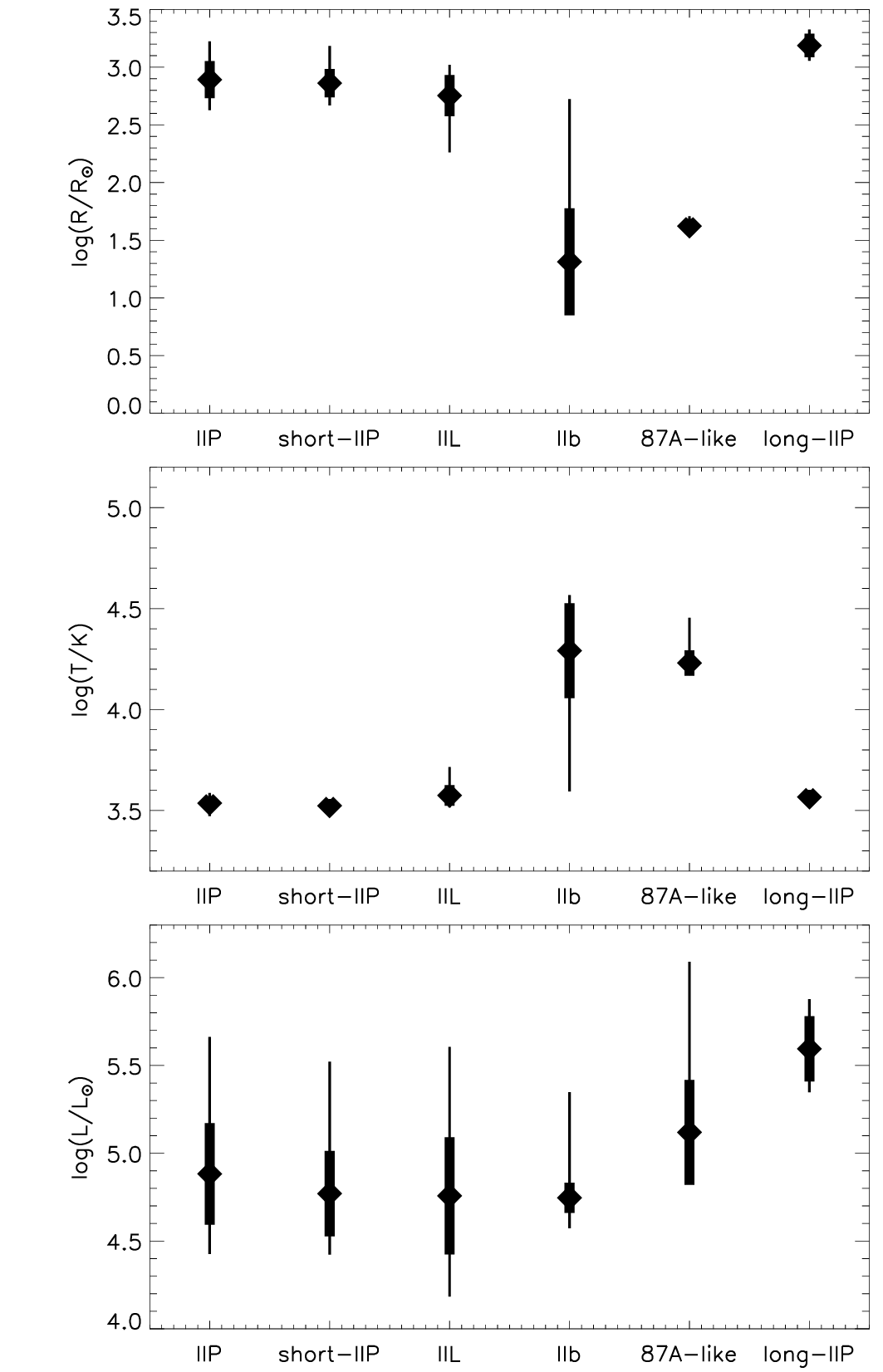}
\includegraphics[width=\columnwidth]{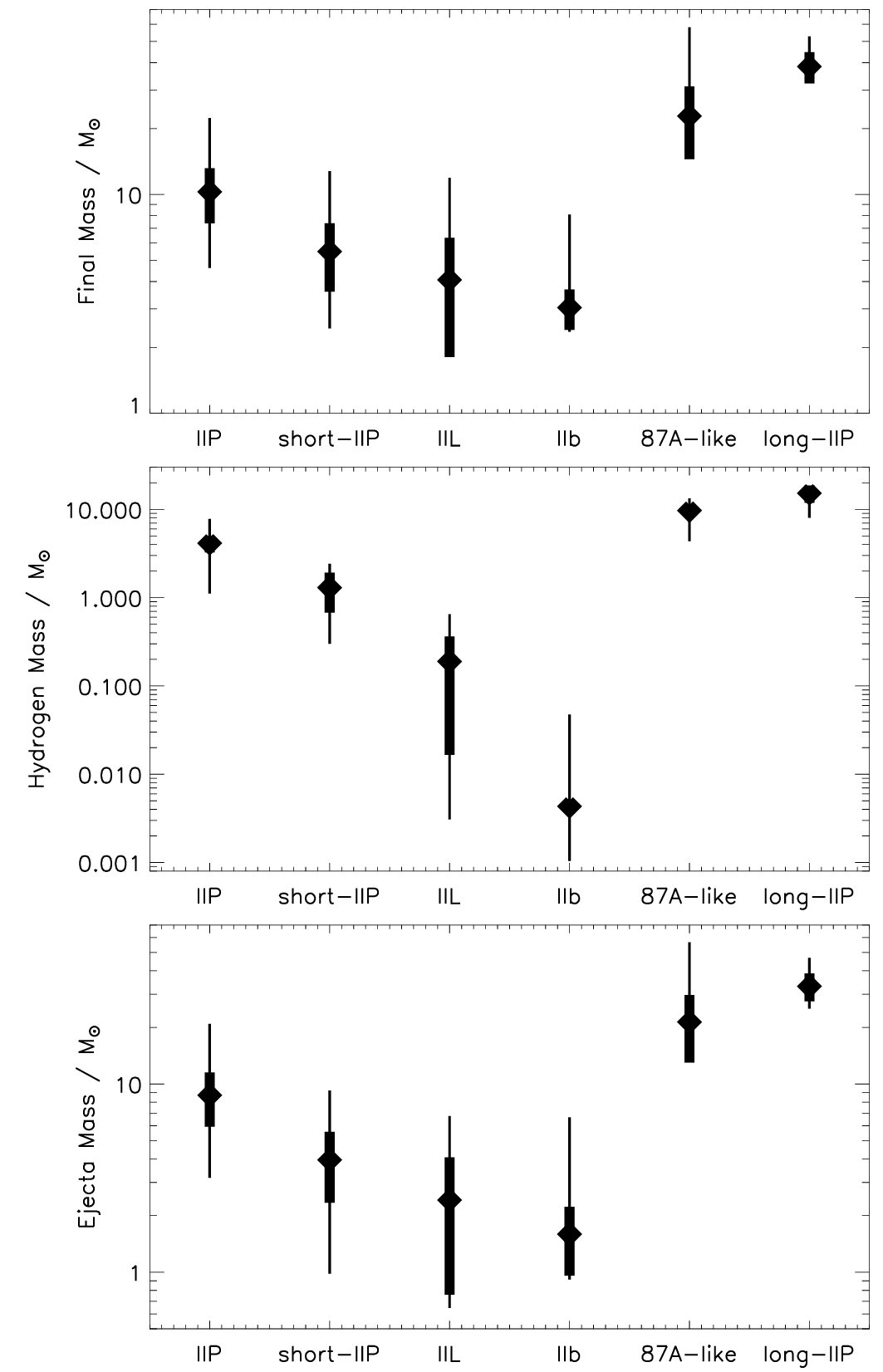}
\caption{The model progenitor parameters of the synthetic lightcurves as typed in Figure 3. Crosses and thick bars indicate the mean and standard deviation of the population, while thin bars indicate the full range spanned by the model set.}\label{Fig4}
\end{center}
\end{figure*}

\begin{figure*}
\begin{center}
\includegraphics[width=\columnwidth]{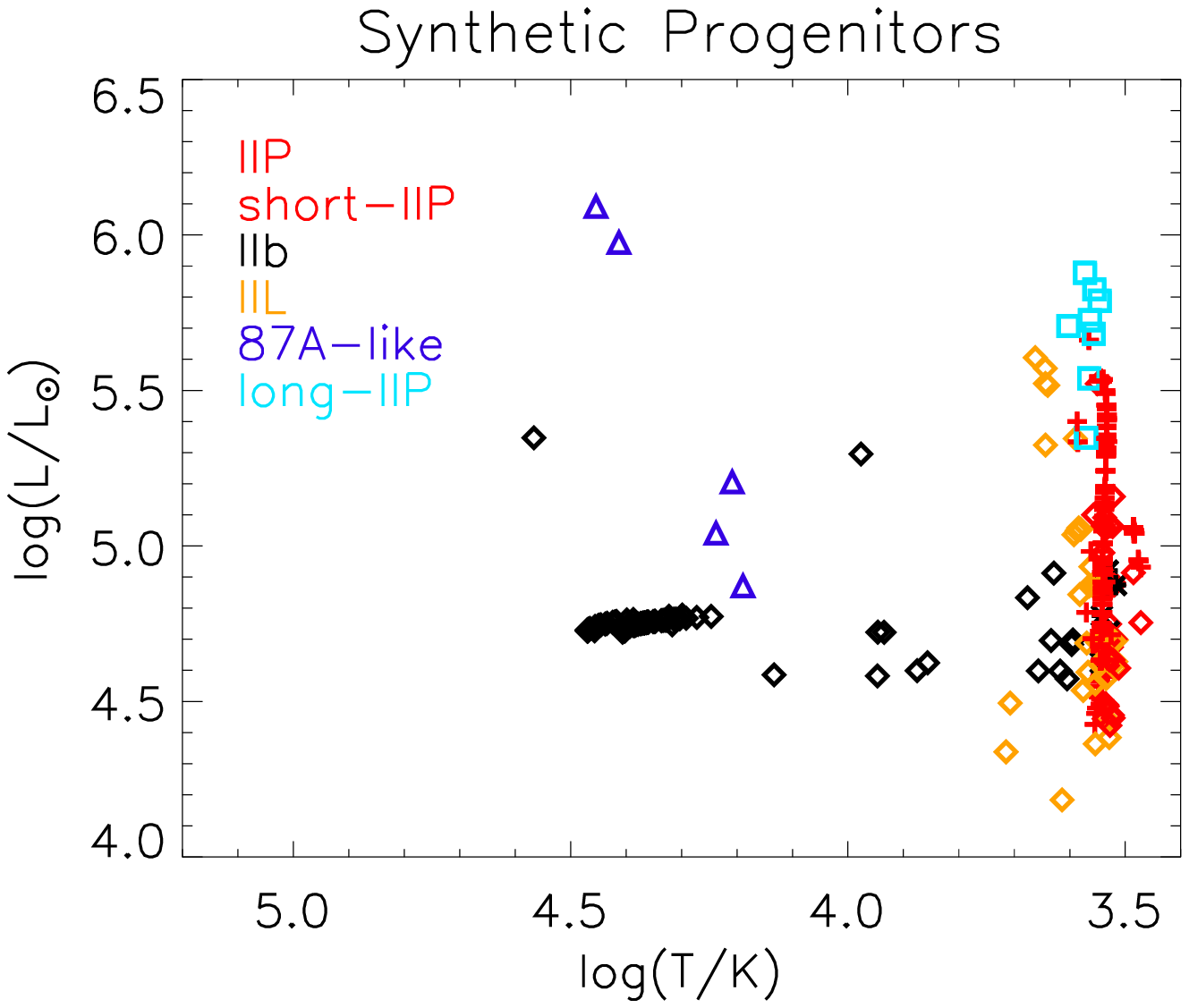}
\includegraphics[width=\columnwidth]{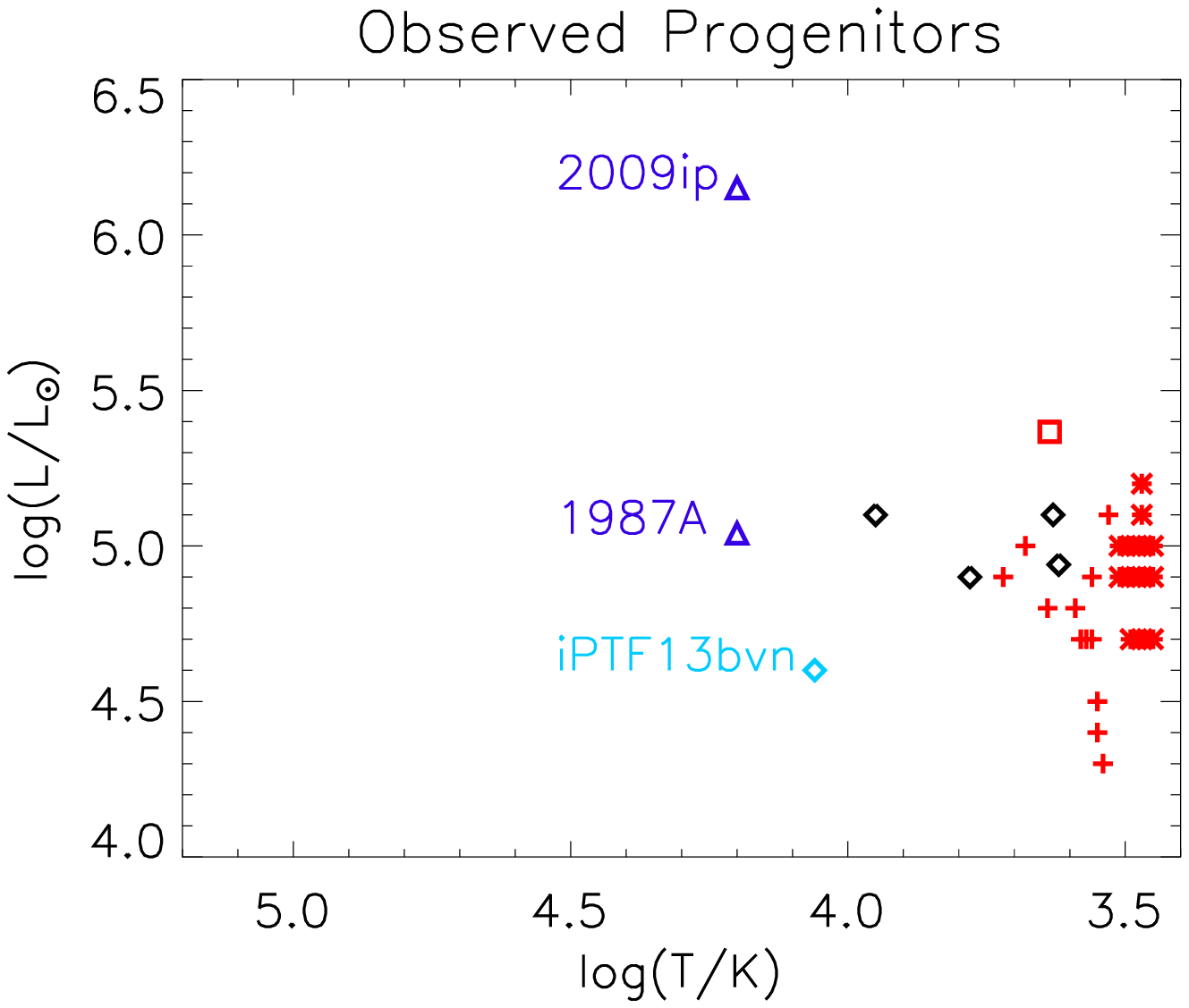}
\caption{Hertzsprung-Russell diagram for our synthetic progenitors (left panel) and for observed progenitors (right panel). In the left panel IIP progenitors are shown as red plus symbols, short-IIP are shown as red diamonds, IIL are shown as orange diamonds, IIb are shown as yellow diamonds, 87A-like progenitors are shown as blue triangles and long-IIP are shown as blue squares. In the right panel these types are supplemented by red asterisks that indicate luminosity upper limits for IIP progenitors, the red square indicates the position of the candidate black-hole forming event \citep[a 'failed' supernova, ][]{2017MNRAS.468.4968A} and the green diamond the progenitor of the stripped envelope supernova iPTF13bvn \citep{2016MNRAS.461L.117E}.}\label{Fig5}
\end{center}
\end{figure*}

\begin{figure*}
\begin{center}
\includegraphics[width=0.65\columnwidth]{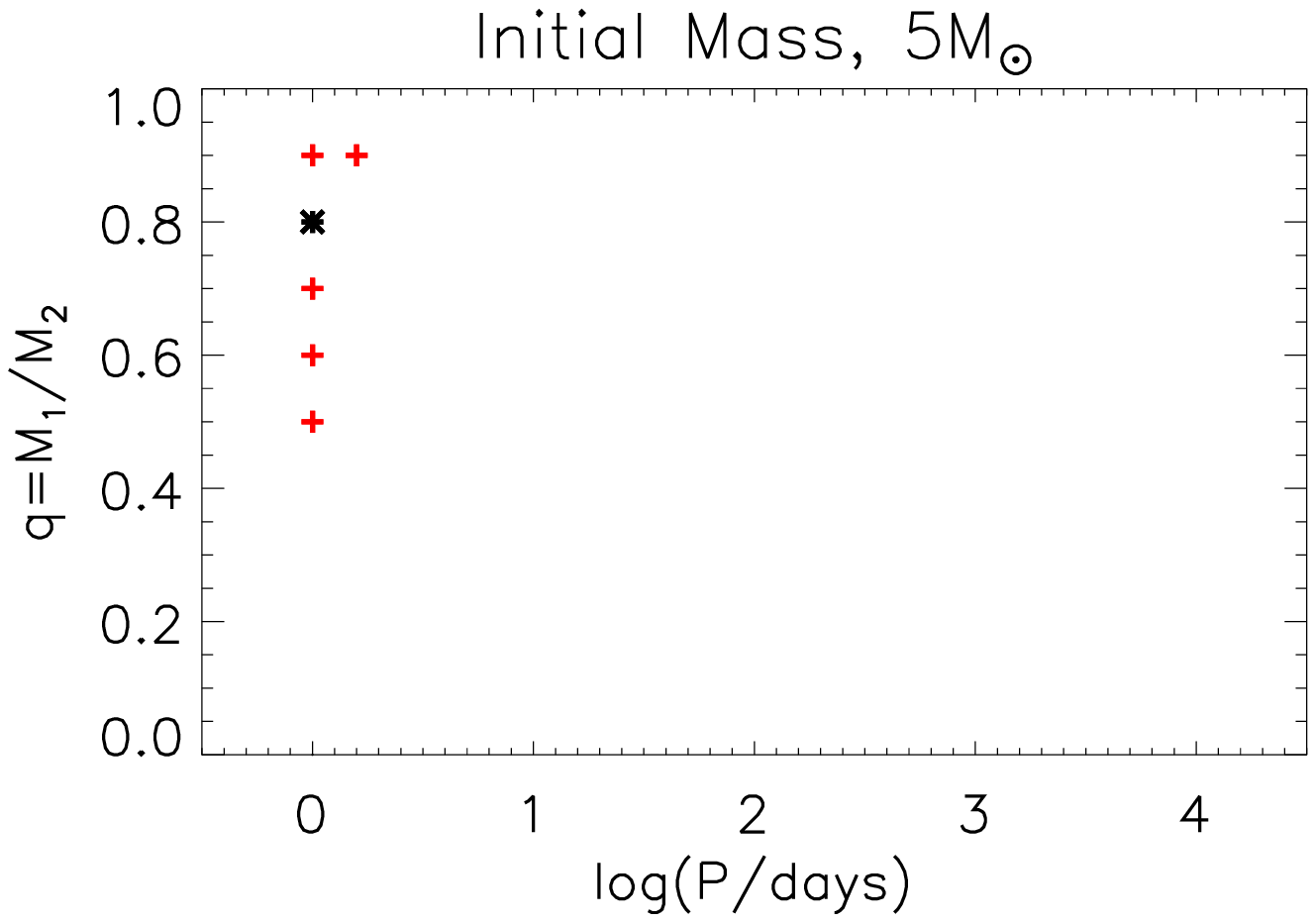}
\includegraphics[width=0.65\columnwidth]{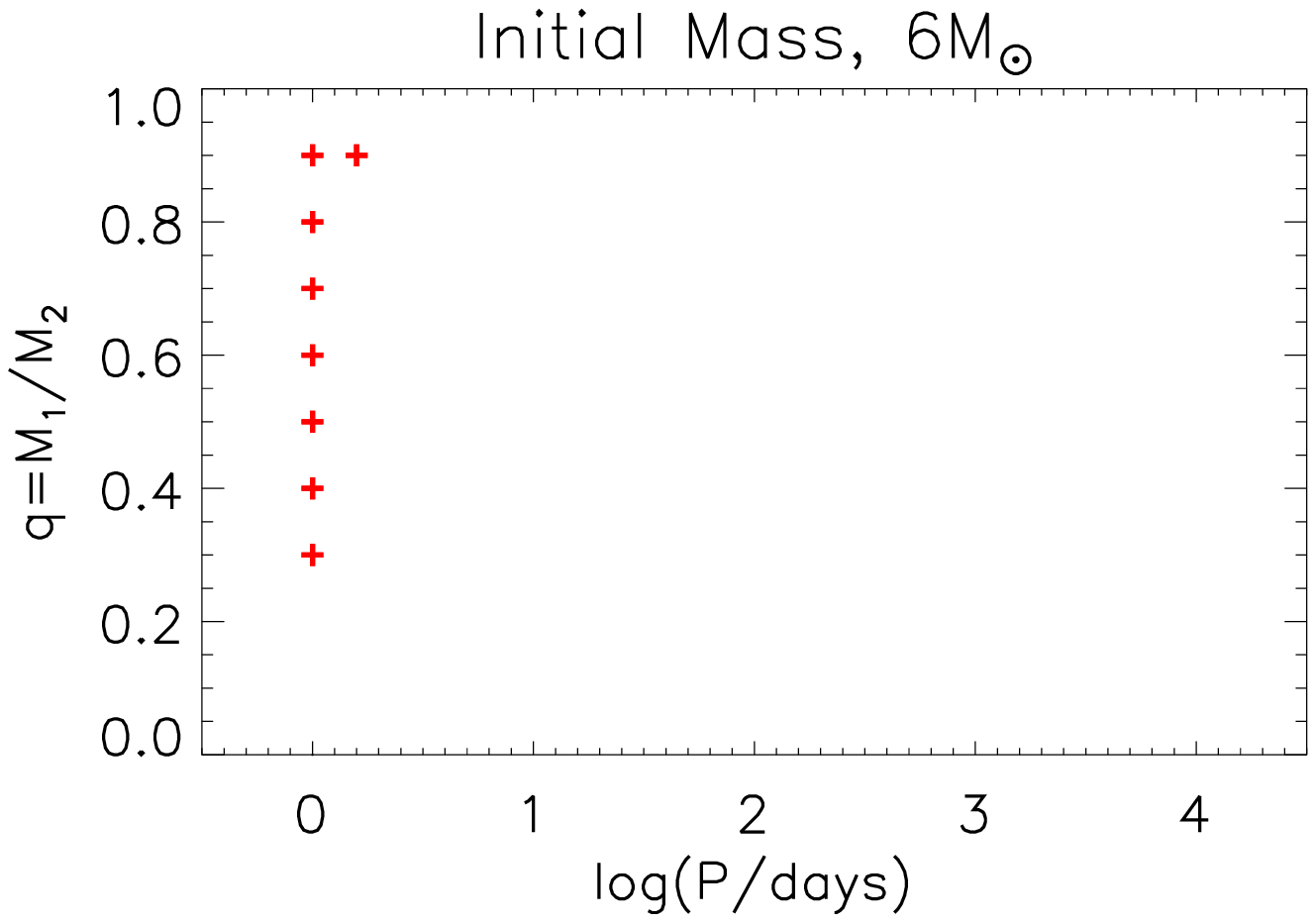}
\includegraphics[width=0.65\columnwidth]{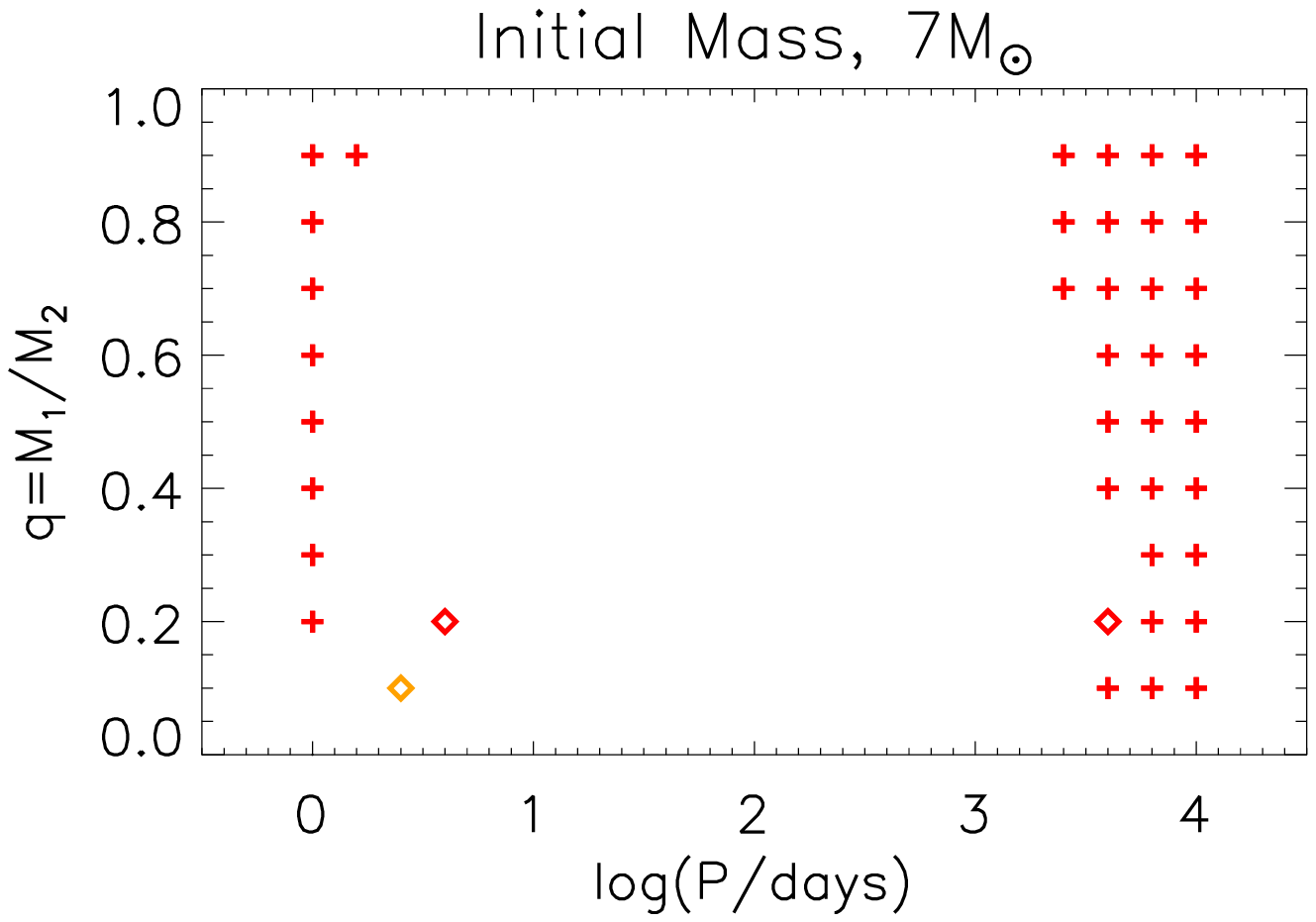}
\includegraphics[width=0.65\columnwidth]{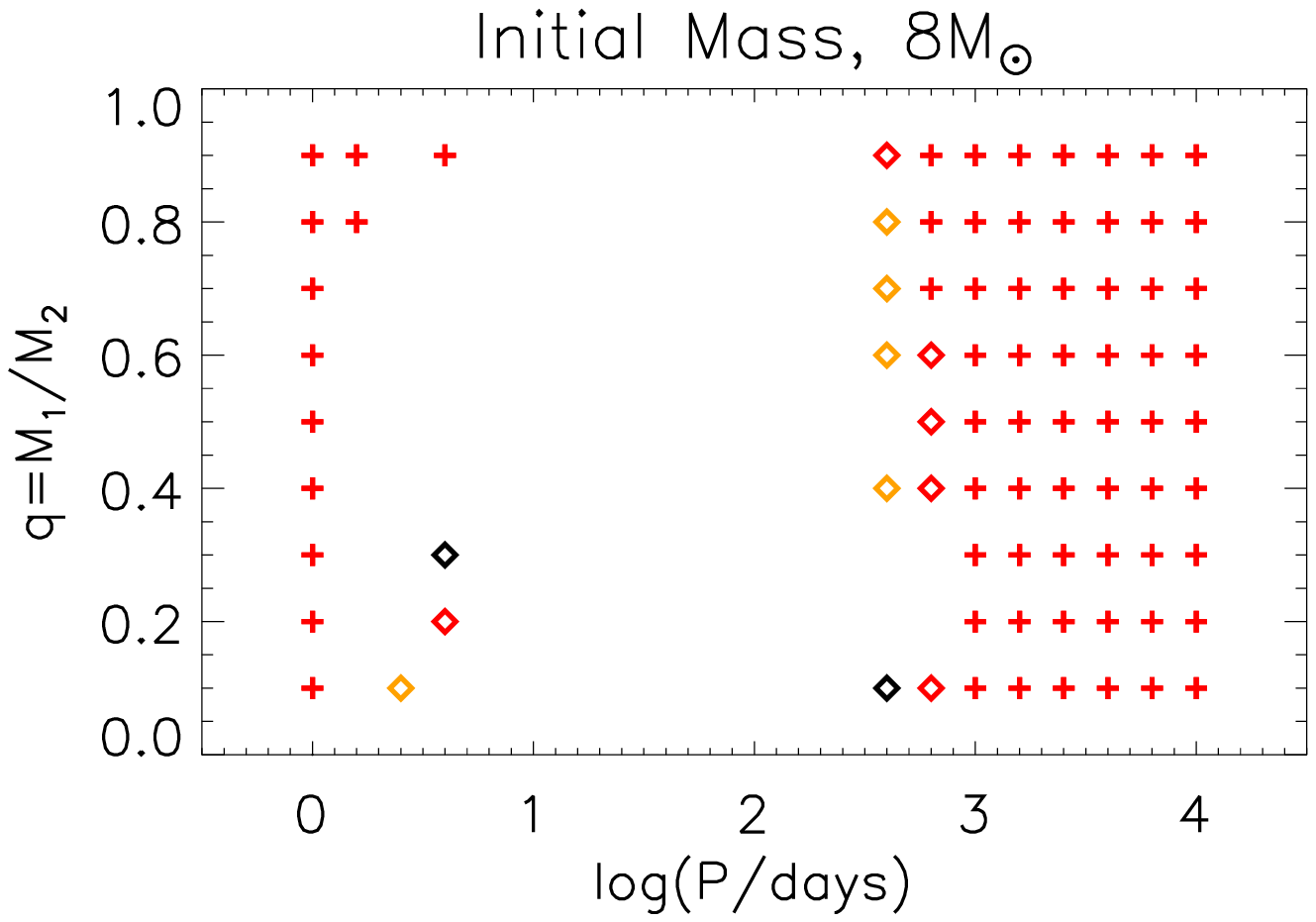}
\includegraphics[width=0.65\columnwidth]{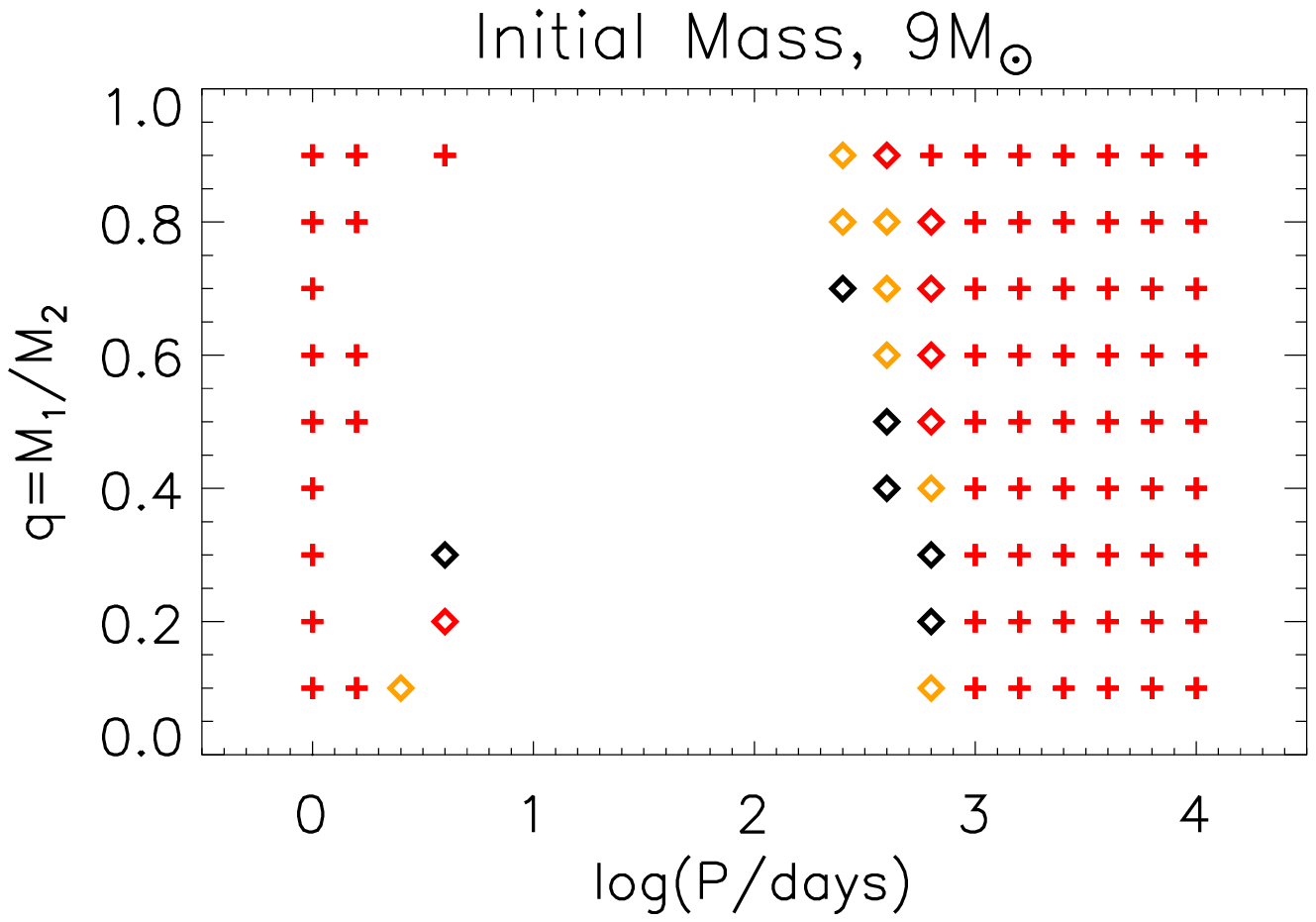}
\includegraphics[width=0.65\columnwidth]{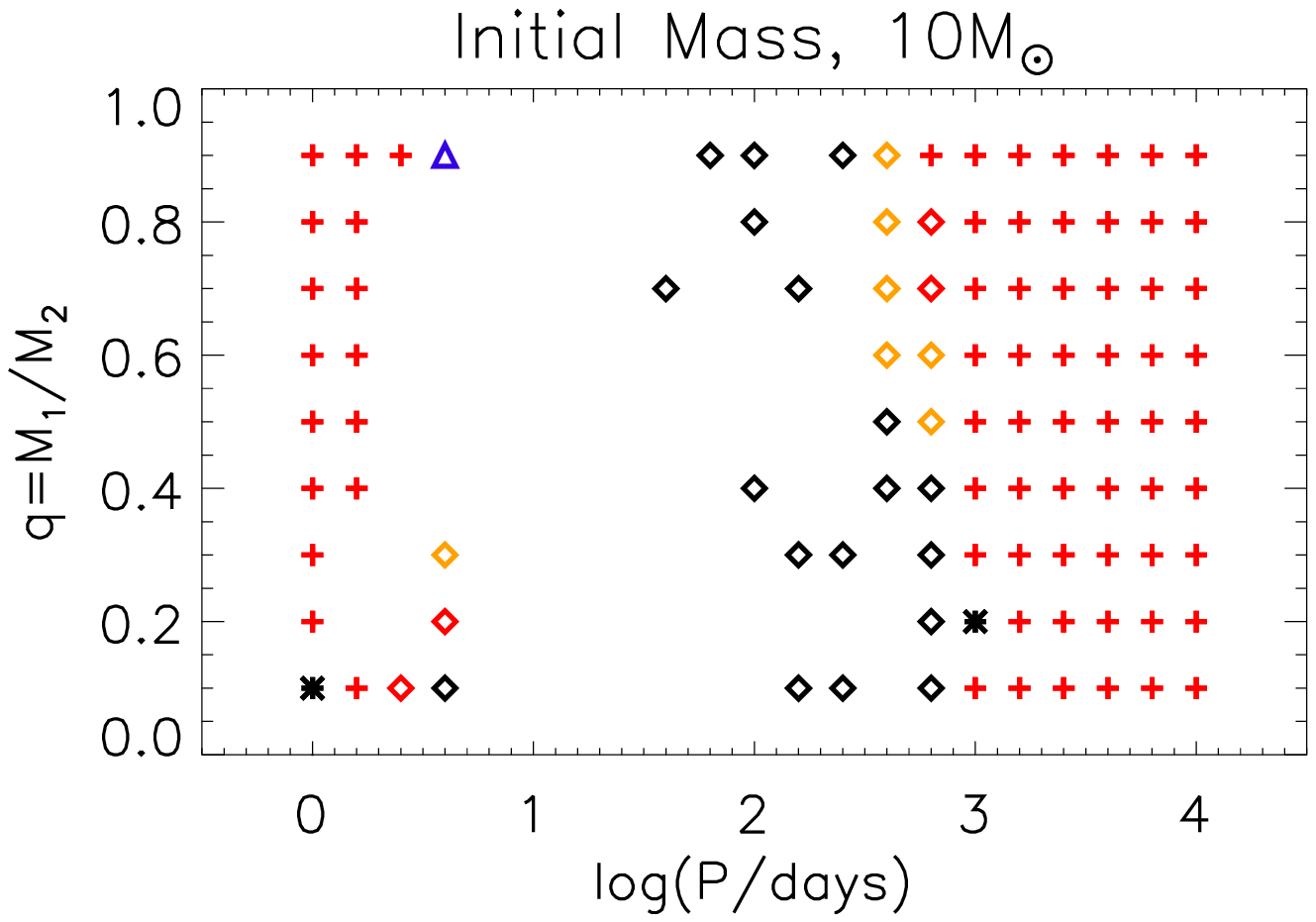}
\includegraphics[width=0.65\columnwidth]{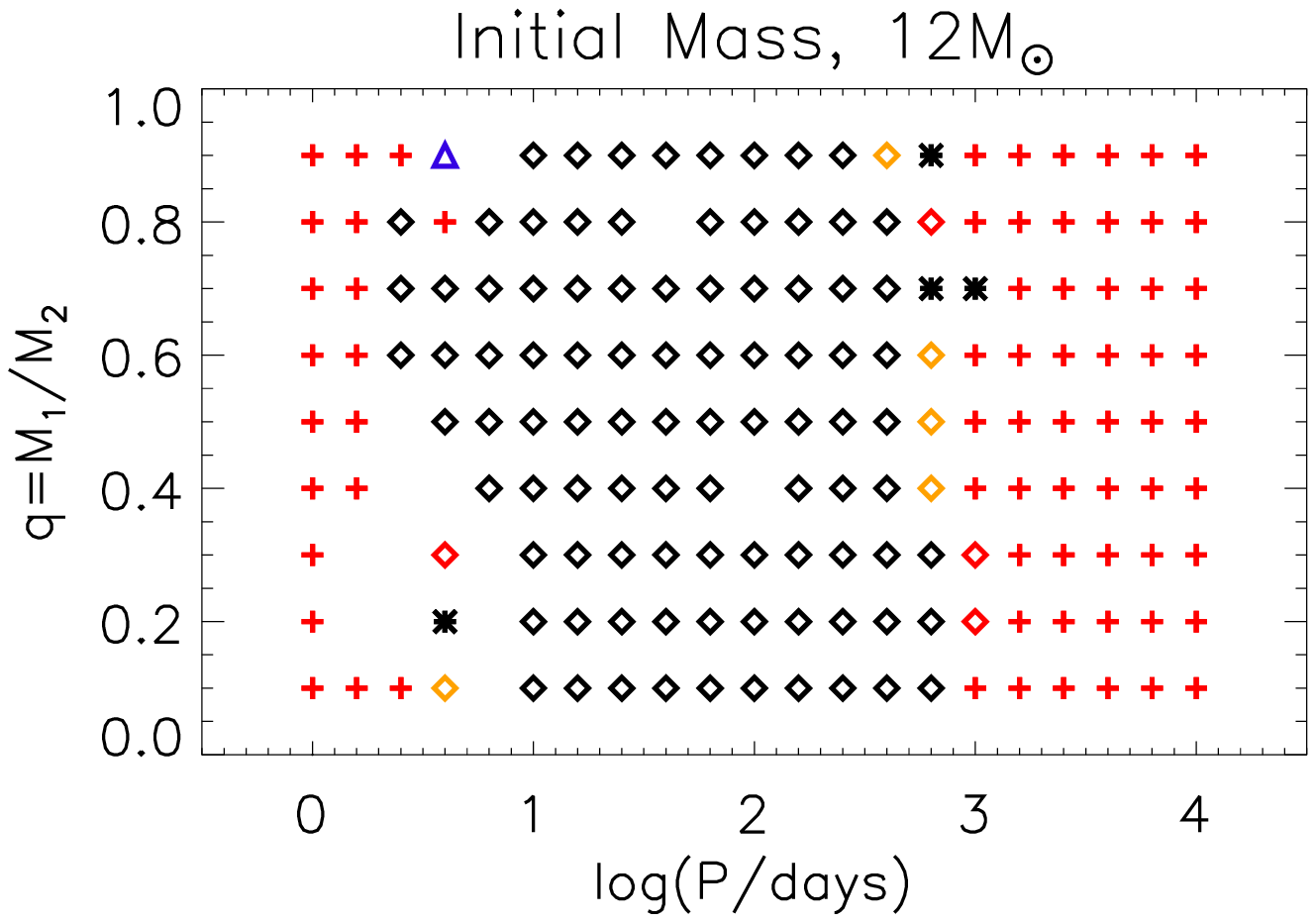}
\includegraphics[width=0.65\columnwidth]{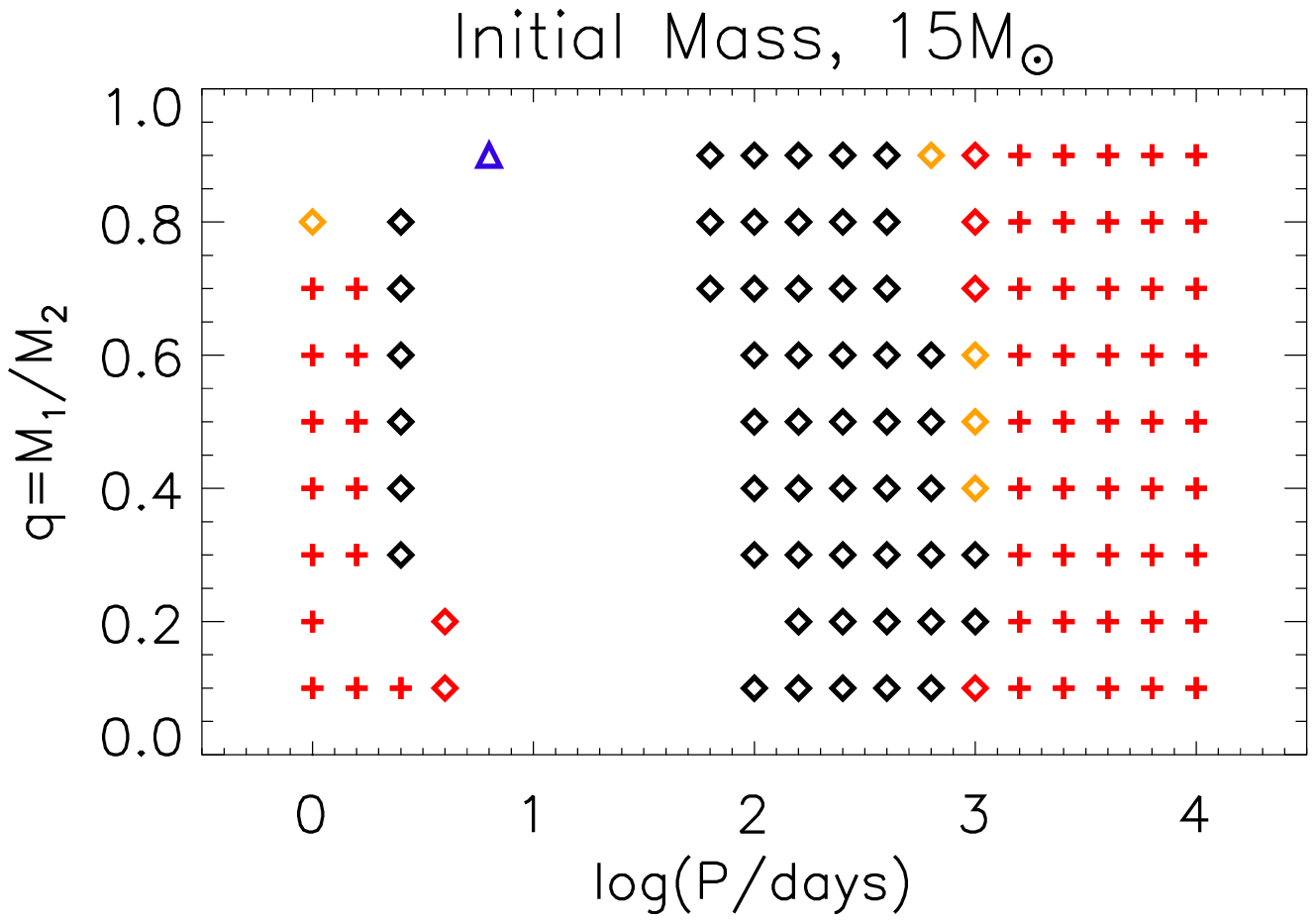}
\includegraphics[width=0.65\columnwidth]{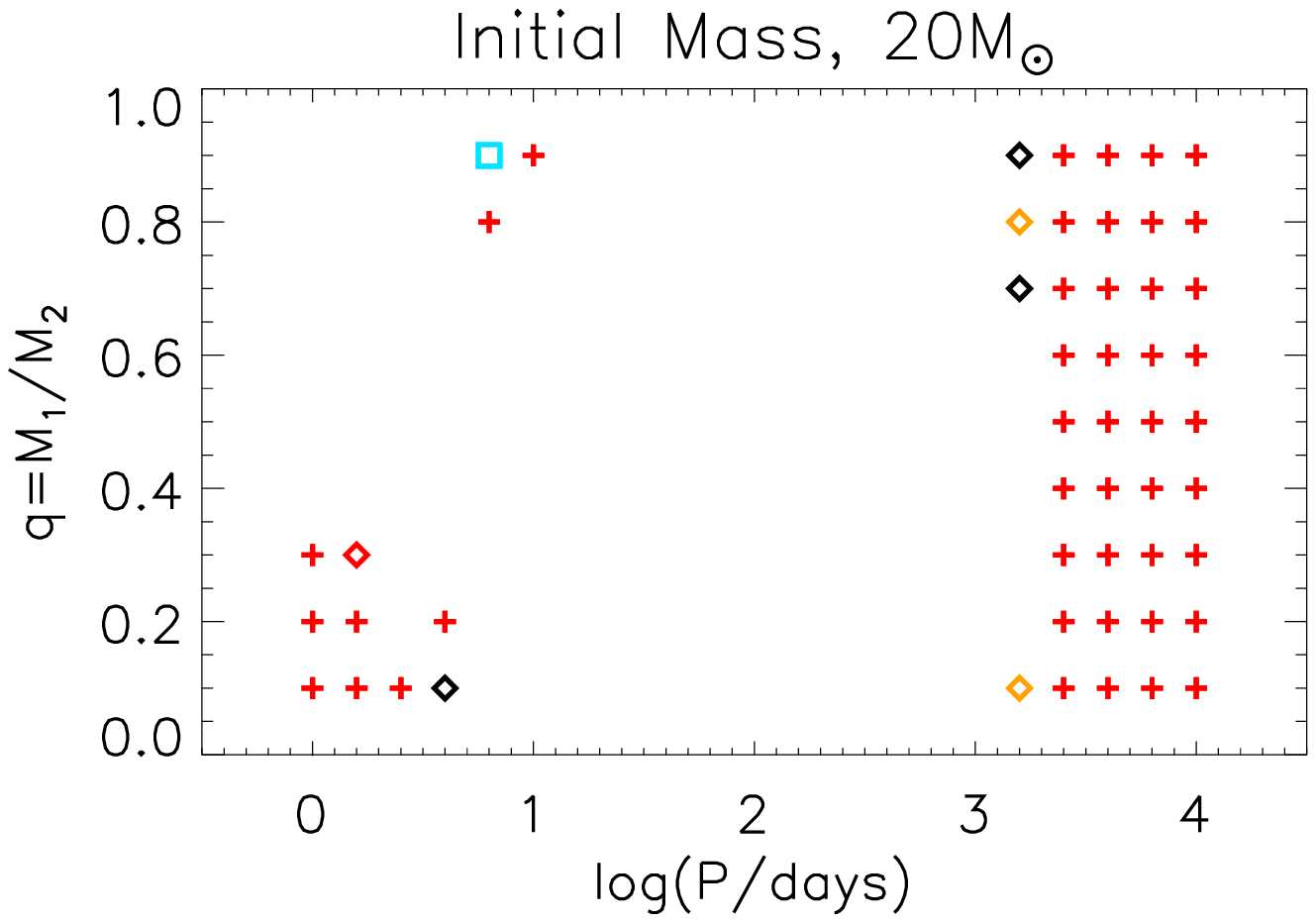}
\includegraphics[width=0.65\columnwidth]{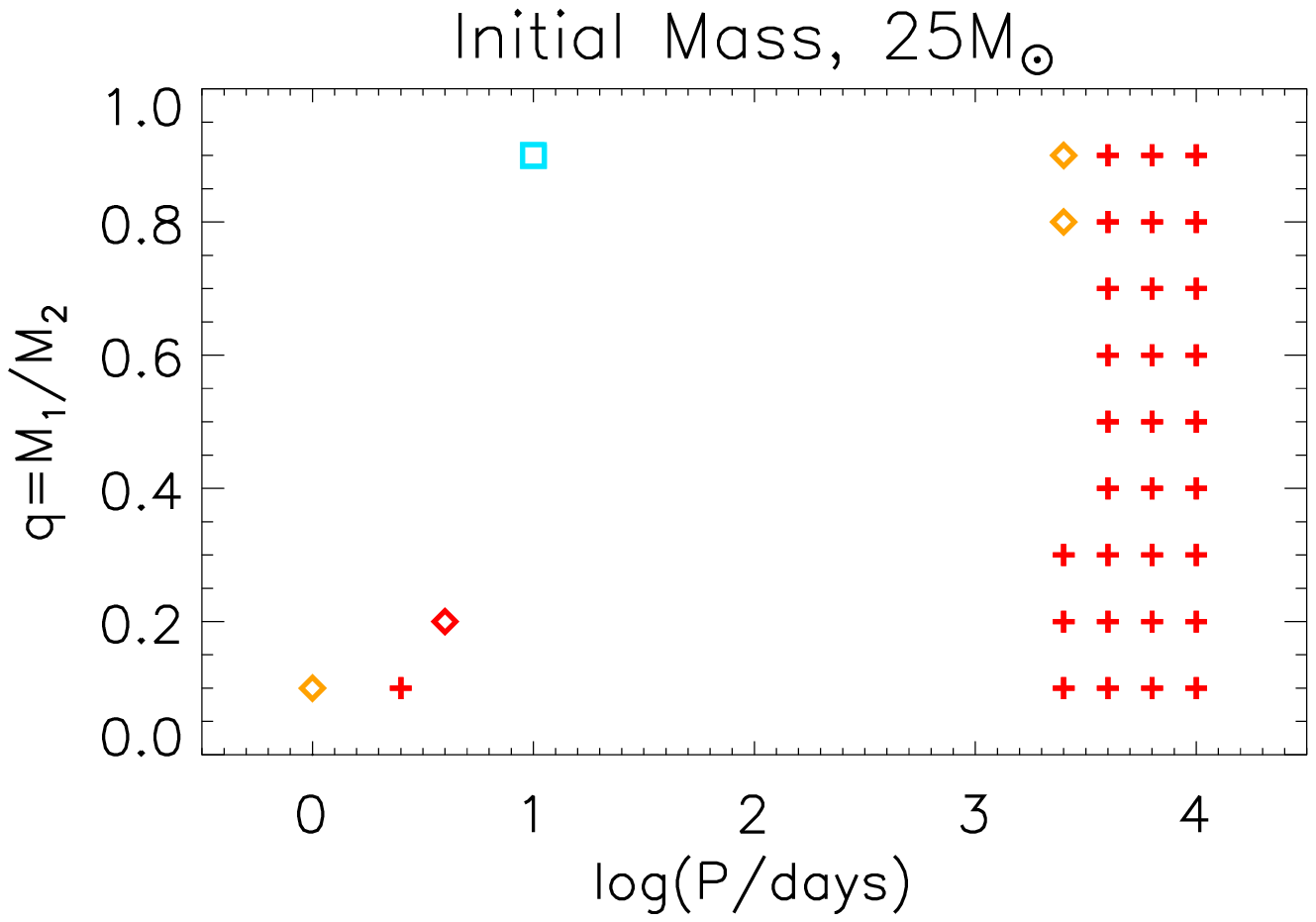}
\includegraphics[width=0.65\columnwidth]{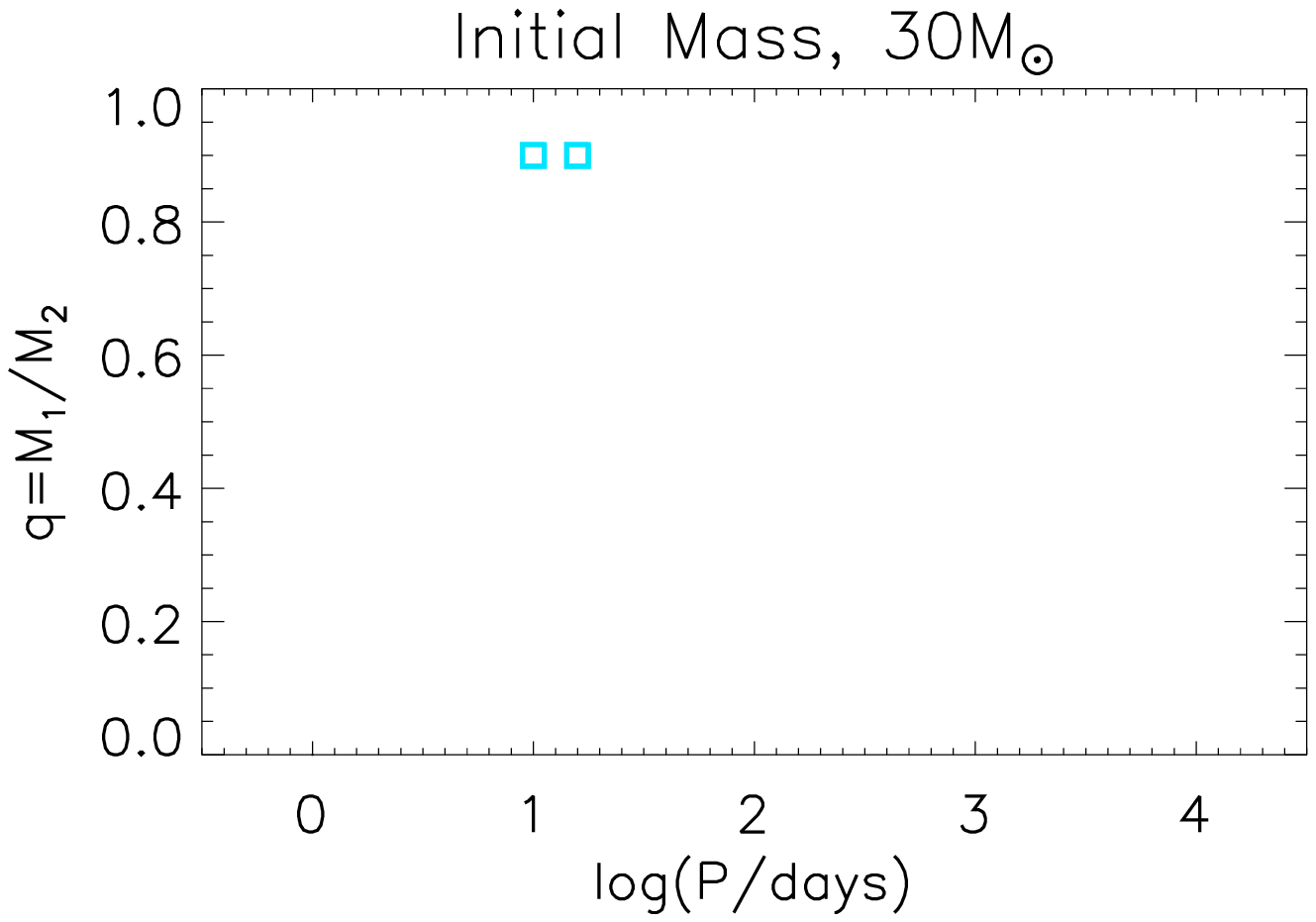}
\includegraphics[width=0.65\columnwidth]{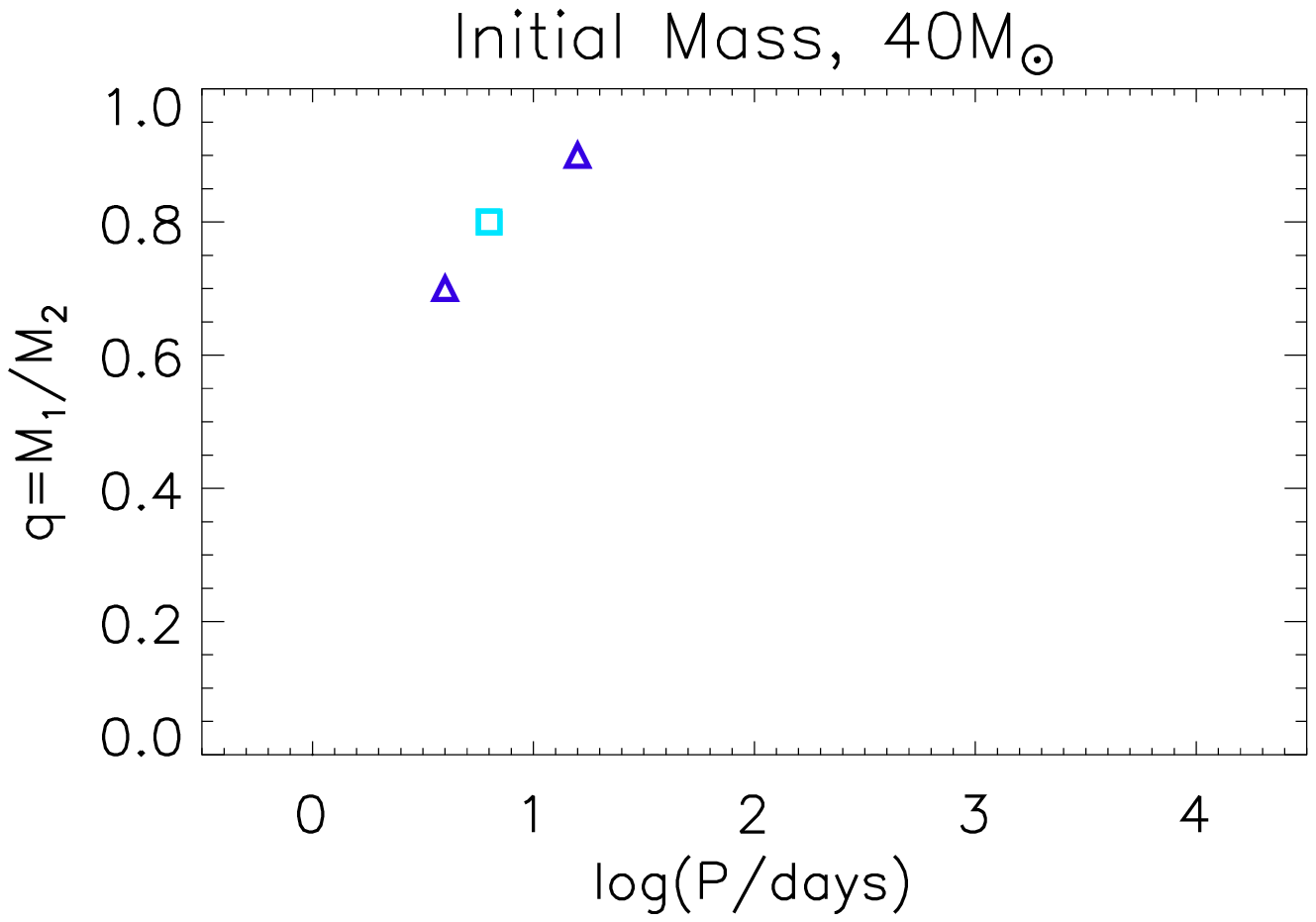}
\caption{The initial binary parameters that lead to different supernova types as a function of initial mass, binary period and mass ratio. Type II supernova sub-classes are coded as indicated in figure \ref{Fig5}. Black asterisks indicates models that did not complete in SNEC and are not included in our analysis.}\label{Fig6}
\end{center}
\end{figure*}

Given that we have now classified the supernova types by their synthetic lightcurves we are able to investigate the nature of the progenitors of these supernova types. Unlike in the case of progenitor parameters derived from observations, we completely know the nature of our model stars. We show our analysis of these progenitors in Figures \ref{Fig4}, \ref{Fig5} and \ref{Fig6}. 

In Figure \ref{Fig4} we summarize the actual physical parameters for the progenitor stars of each supernova sub-class. For each parameter and supernova type we show the mean, standard deviation and range. Most of the type II subclasses here show very similar progenitor radius and temperature distributions with only the IIb and 87A-like supernovae deviating to (on average) smaller and hotter progenitors. By contrast each supernova type shows a range of progenitor luminosities, although those with the highest luminosities are the 87A-like and long-IIP supernovae which begins to indicate their nature may be arising from more massive progenitors.

The reason for these differences in the surface parameters of the progenitor stars are indicated by the panels in Figure \ref{Fig4} that show the final mass, hydrogen mass and ejecta mass of the progenitors. There is a decreasing trend of these masses in the sequence from IIP, short-IIP, IIL to IIb. We note that many of our IIb have vanishingly small amounts of hydrogen so may be more likely to be observed as Ib, hydrogen-free SNe. However this does support the long proposed link of supernova type to ejecta masses \citep{1995NYASA.759..360N,1996IAUS..165..119N,2011MNRAS.414.2985D,2016MNRAS.458.1618D}.

The masses for the 87A-like and long-IIP supernovae are significantly greater than the remaining events, due to them arising from binary systems that have experienced a merger. The long-IIP events require the most massive progenitors. Such massive stars are known to throw off substantial amounts of material well before their explosive detonation, leading to a dense and irregular circumstellar medium. This suggests such events may appear very different if observed in nature - possibly as type IIn supernovae. Future modelling of such events will therefore need to fully account for the circumstellar medium around the supernova progenitors.

An alternate way of displaying the information in Figure \ref{Fig4} is to plot the progenitor stars on a Hertzsprung-Russell diagram as shown in Figure \ref{Fig5}. The type IIP progenitors are all red supergiants, as are the IIL progenitors. Type IIL progenitors however are scattered to warmer temperatures. The type IIb progenitors extend out to the hottest temperatures and may be observed as Wolf-Rayet stars in pre-explosion images. The 87A-like progenitors are all blue supergiants and the long-IIP progenitors detonate at the very tip of the red supergiant branch. Such progenitors only exist as type II events due to late time mergers adding hydrogen back to the evolved progenitor stars. Whether these very luminous red supergiants and the very luminous 87A-like progenitors exist or will depend the uncertain mass-loss rates of these very massive stars. 

In the second panel of Figure \ref{Fig5} we provide an equivalent diagram showing the inferred properties of supernova progenitors observed in pre-explosion imaging, as collated in \citet{2017PASA...34...58E}. The observed type IIP progenitors match our predictions well. Supernova 1987A lies in the same place as our synthetic progenitors for events with similar lightcurves. However the observed IIb progenitors tend to be concentrated to cooler temperatures than expected for synthetic progenitors. In addition the observed progenitor of iPTF13bvn (a stripped envelope type Ib event) matches our proposed IIb progenitors. In not producing synthetic spectra of our SNe, we are uncertain which supernovae might be observed as Ib or IIb and thus apply a somewhat arbitrary threshold based on surface hydrogen abundance, which represents a limitation of this study. We could potentially increase the minimum mass of hydrogen required in our progenitor models for classification as IIb SNe, but we leave our predictions as they are here to demonstrate that there may be a continuum of possible supernovae between type IIb to Ib. 

A further insight into the progenitors of the different type of supernovae is provided by how the different supernovae types arise from binaries of different initial period and mass ratio at each initial mass. In Figure \ref{Fig6} we show the outcome for the primary star in each binary as a function of mass, initial period and mass ratio, using the same symbols as in Figure \ref{Fig5} to indicate supernova type. 

At all masses the closest binaries, with initial periods of days, always give rise to type IIP supernovae as the stars effectively evolve as single stars after an early merger. The  binaries with initial periods of around 1000 days are again so wide that the primary star's evolution is effectively that of a single star, and again these explode in type IIP SNe. For binaries just below this limit, different supernova types occur depending on the amount of mass lost in interactions. However between 7 and 10\,M$_{\odot}$ there are large initial period ranges where no supernova occurs. Here the interactions are so strong that the stars evolve to become white dwarfs rather than SNe. Only above 12\,M$_{\odot}$ do all stars again produce SNe. At the highest masses, gaps in the panels indicate parameter ranges when hydrogen-free type Ib/c supernovae can be expected to occur.

An interesting conclusion to draw from this is that most type IIb supernovae arise from progenitor stars in the mass range 10 to 15\,M$_{\odot}$. Another is that supernovae which result from stellar mergers close to core-collapse are relatively rare and arise from a very small range of initial binary parameters.

\section{Discussion and conclusions}

In this article we have undertaken, to our knowledge, the first attempt at a supernova lightcurve population synthesis, exploding a significant number of supernova progenitor models rather than only a few stars at a time. This allows us to gain insight into the range of possible supernova that arise from a population of stars, and also to examine in detail how the parameters such as radius, mass and internal structure affect the resultant supernova lightcurves.

It is important to outline the caveats and limitations of this approach. We have only exploded a limited number of our approximately 100,000 possible supernova progenitor models due to the constraints of computational resources. We have concentrated on a reduced but representative grid of initial masses and have only considered the primary stars of the binaries. In many cases, particularly in binaries with a mass ratio close to unity, the surviving companion may also go supernova at a later time. We have also explored only one choice of explosion energy, nickel mass and strength of nickel mixing. This has the advantage that the diversity of behaviour is attributable only to the stellar structure and the initial parameters determining it. It likely means that our prediction of the diversity of supernova lightcurves is an underestimate.

The assumption of a constant explosion energy is too simple and much greater diversity would be obtainable by increasing or decreasing our assumed $10^{51}\, {\rm erg\,s^{-1}}$. There is strong evidence from observations that this is the case \citep{2013MNRAS.436.3224P,2015ApJ...806..225P}. Then to complicate matters further this varying explosion energy may also vary the mass of nickel or mass of the remnant formed in the supernova and whether an observable event is possible at all \citep{2016ApJ...818..124E,2016ApJ...821...38S,2018ApJ...860...93S}. We will explore some variation in explosion energy in future work, although given the large range of models, this remains computationally challenging. Finally there may also be central-engines such as magnetars that continue to inject energy after the stellar explosion and thus can provide lightcurves that look like those that are observed \citep{2016ApJ...833...64M,2017MNRAS.472..224S}.

We have also not included the circumstellar medium around the progenitor in terms of its pre-supernova stellar wind in our models. \citet{2018MNRAS.476.2840M} and \cite{2018ApJ...858...15M} have recently shown that this can significantly change the early-time lightcurve of the supernovae. Additionally the metallicity of the progenitor stars will change the interior structure of the star and also the density of the pre-supernovae wind - thus both of these aspects should be included in future more detailed studies.

Given these caveats we can draw the following conclusions:
\begin{enumerate}
\item Much of the diversity of type II supernova lightcurves arises from the effects of binary interactions on stellar structure. Single-star progenitors of the same mass range all produce type IIP supernovae, only the various degrees of mass loss from binaries in different strengths of interactions give different envelope masses and thus different lightcurve shapes.
\item Stellar mergers can lead to mass gain and provide a quite different lightcurve shape similar to supernova 1987A or very long-IIP like supernovae. For the latter, their higher masses suggest their explosion parameters may be different and so such events may look different if observed in nature.
\item There is a progression in how much mass is lost in progenitor stars, with IIP supernovae having the most ejecta mass, with decreasing ejecta mass over the types IIL to IIb.
\item The expected location of the synthetic progenitors in the HR diagram from observed supernova progenitors are consistent.
\item Different supernova types come from very specific initial binary parameters. 
\end{enumerate}

The final most important result from this study is showing the power of supernova population synthesis. Rather than just deciding the supernova type of a stellar model by using parameters such as hydrogen or ejecta mass we can make firmer predictions on the link between stellar death throes in supernovae and the nature of their progenitor stars.

\begin{acknowledgements}
This work would not have been possible without use of the
NeSI Pan Cluster, part of the NeSI high-performance computing
facilities. New Zealand's national facilities are provided
by the NZ eScience Infrastructure and funded jointly
by NeSI's collaborator institutions and through the Ministry
of Business, Innovation \& Employment's Research Infrastructure programme. URL: \texttt{https://www.nesi.org.nz.}.
JJE acknowledges support from the
University of Auckland and thanks the OCIW Distinguished Visitor Program at the Carnegie Observatories, which funded JJE's visit to be gain advice from Anthony L. Piro on using SNEC.
ERS acknowledges support from the University of Warwick. LX acknowledge the grant from the National Key R\&D Program of China (2016YFA0400702), the National Natural Science Foundation of China (No. 11673020 and No. 11421303), and the National Thousand Young Talents Program of China. NR and NYG both acknowledges support from University of Auckland's Summer Scholarships.

\end{acknowledgements}

\bibliographystyle{mnras}

\end{document}